%% file: paper.tex
\documentclass[fleqn,10pt]{wlscirep}
\usepackage[utf8]{inputenc}
\usepackage[T1]{fontenc}
\usepackage{hyperref}
\usepackage{orcidlink}
\usepackage{subcaption}
\usepackage[group-separator={,}]{siunitx}
\usepackage{graphicx}
\DeclareGraphicsExtensions{.pdf,.png}
\graphicspath{ {./figures/} }

\usepackage{authblk}

\usepackage{xurl}

\usepackage{lineno}

\usepackage{soul}

\usepackage{glossaries}

\input{acronyms.tex}

\usepackage[numbers]{natbib}

\input{population_correlations}
\input{population_correlations_unstretched}

\input{toronto_stats}

\usepackage{xintexpr}
\def\includecomplexity{1}

\input{fix_wlscirep_sectionref}

\title{Revealing urban area from mobile positioning data}

\author[1,*]{Gergő Pintér \orcidlink{0000-0003-4731-3816}}
\affil[1]{ANETI Lab, Corvinus Institute for Advanced Studies, Corvinus University of Budapest, Budapest, 1093, Hungary}

\affil[*]{Corresponding author: \href{mailto:gergo.pinter@uni-corvinus.hu}{gergo.pinter@uni-corvinus.hu}}

\keywords{mobile positioning data, urban mobility, YJMob100K, HuMob2023 challenge, reverse-engineering}

\begin{abstract}
Researchers face the trade-off between publishing mobility data along with their papers while simultaneously protecting the privacy of the individuals.
In addition to the fundamental anonymization process, other techniques, such as spatial discretization and, in certain cases, location concealing or complete removal, are applied to achieve these dual objectives.
The primary research question is whether concealing the observation area is an adequate form of protection or whether human mobility patterns in urban areas are inherently revealing of location.
The characteristics of the mobility data, such as the number of activity records or the number of unique users in a given spatial unit, reveal the silhouette of the urban landscape, which can be used to infer the identity of the city in question.
It was demonstrated that even without disclosing the exact location, the patterns of human mobility can still reveal the urban area from which the data was collected.
The presented locating method was tested on other cities using different open data sets and against coarser spatial discretization units.
While publishing mobility data is essential for research, it was demonstrated that concealing the observation area is insufficient to prevent the identification of the urban area.
Furthermore, using larger discretization units alone is an ineffective solution to the problem of the observation area re-identification.
Instead of obscuring the observation area, noise should be added to the trajectories to prevent user identification.
\end{abstract}

\begin{document}

\flushbottom
\maketitle

\thispagestyle{empty}


\section*{Introduction}
\label{sec:introduction}

In order to ensure the reproducibility of research findings, it is advisable for scholars to publish both data and code alongside their papers.
Nevertheless, the publication of mobility data raises issues regarding privacy, as the process of anonymization is complex and challenging.
It has been demonstrated on numerous occasions that the implemented anonymization process was inadequate.
The users in the Netflix Prize dataset were re-identified using \acrfull*{IMDb} as a source of background knowledge \cite{narayanan2008robust}.
The taxi IDs of the \acrfull*{NYC} taxi data were also re-identified, revealing the drivers' identity due to poor anonymization technique \cite{douriez2016anonymizing}.
In combination with paparazzi photographs from social media, which often capture the unique IDs of the cabs, celebrities can be tracked in the \acrshort*{NYC} dataset \cite{tayouri2016social}.
Similarly, the ticket IDs in the published data of the Riga public transport were easily de-anonymized \cite{lavrenovs2016privacy}, revealing the ticket type, which could lead to privacy attacks.

Mobile positioning data can be affected as well.
As demonstrated by Sharad and Danezis in their analysis of the anonymization of the \acrfull*{D4D} challenge data \cite{blondel2012data}, was found to be inadequate \cite{sharad2013anonymizing}.
Even if the anonymization process was performed correctly, and the user is not identifiable by any columns, the visited locations still provide some attack vectors based on the individuals' behavior.
De Montjoye et al. showed that the vast majority of the individuals could be identified by only four spatial data points \cite{de2013unique}.
Another prevalent technique is spatial discretization \cite{faraji2023point2hex,bergroth202224,xu2021towards}, which can limit the accuracy of geographic locations to a certain level.

This paper presents evidence that concealing the observation area of mobility data is an ineffective solution to the privacy issue.
To achieve this, I primarily use the `YJMob100K' data set \cite{yabe2024yjmob100k}, a metropolitan-scale, longitudinal, anonymized mobility trajectory data set that aims to serve as a benchmark dataset of human mobility \cite{yabe2024enhancing}.
The data provider is Yahoo Japan Corporation, and the data follows \num{100000} individuals across a 90-day period in an undisclosed, highly populated metropolitan area in Japan \cite{yabe2024yjmob100k}.
In this data set the geographic locations of the individuals are discretized (into 500-meter by 500-meter cells), and the location of the observation area is undisclosed.
Furthermore, the precise dates are not provided; instead, the days are numbered relatively (e.g., day 1, day 2, etc., up to day 90) and the time is also discretized into 30-minute intervals.

The data has been discretized in multiple dimensions, and user IDs are sequential numbers.
However, the main question is whether the undisclosed observation area provides sufficient protection.
Alternatively, the characteristics of human mobility can reveal in which urban area the mobility data was captured.
Despite the absence of geographic locations for the spatial grid cells, the activity distribution within the grid can be used to infer the city landscape.
Once the urban area was deduced using a map, a template matching method was applied to find the exact grid location on the map.
This process enables the reconstruction of the spatial grid used to discretize the mobility data.

The presented approach was generalized in two directions.
Initially, coarser discretization resolutions were simulated, demonstrating that city characteristics cannot be easily suppressed.
Second, the location technique, based on template matching, was presented for four other cities from three additional data sets.
Furthermore, in one of these cases, H3 hexagons were used instead of a grid, indicating that the discretization scheme is also irrelevant.

The majority of human activity occurs in urban environments.
Consequently, when a sufficient amount of location data is available, it will inherently reveal the city in which the data was captured.
The bottom line is that privacy by obscurity is not viable solution for ensuring true privacy.
Previous studies have shown that adding noise to the location trajectories can mitigate privacy issues \cite{acs2021privacy,douriez2016anonymizing,mir2013dp}, when synthetic mobility data is not a viable option.

\section*{Results}
\label{sec:results}

The urban area in which the activities were recorded was determined by analyzing the visible silhouette of the urban landscape in the activity heatmap \cite[Figure~6]{yabe2024yjmob100k}, and assuming that the low-activity areas are partly water surfaces.
The urban area was identified as the Nagoya metropolitan area including Mikawa Bay and Ise Bay.
Subsequently, the spatial grid, which was used to discretize the geographic locations of the individuals, was reconstructed.
This section presents a mobility analysis, which serves as a validation of the grid reconstruction.
The reverse-engineering process is detailed in the \nameref*{sec:methods} section.

\begin{figure}[th]
    \centering
    \begin{subfigure}[t]{0.245\linewidth}
        \includegraphics[width=\linewidth]{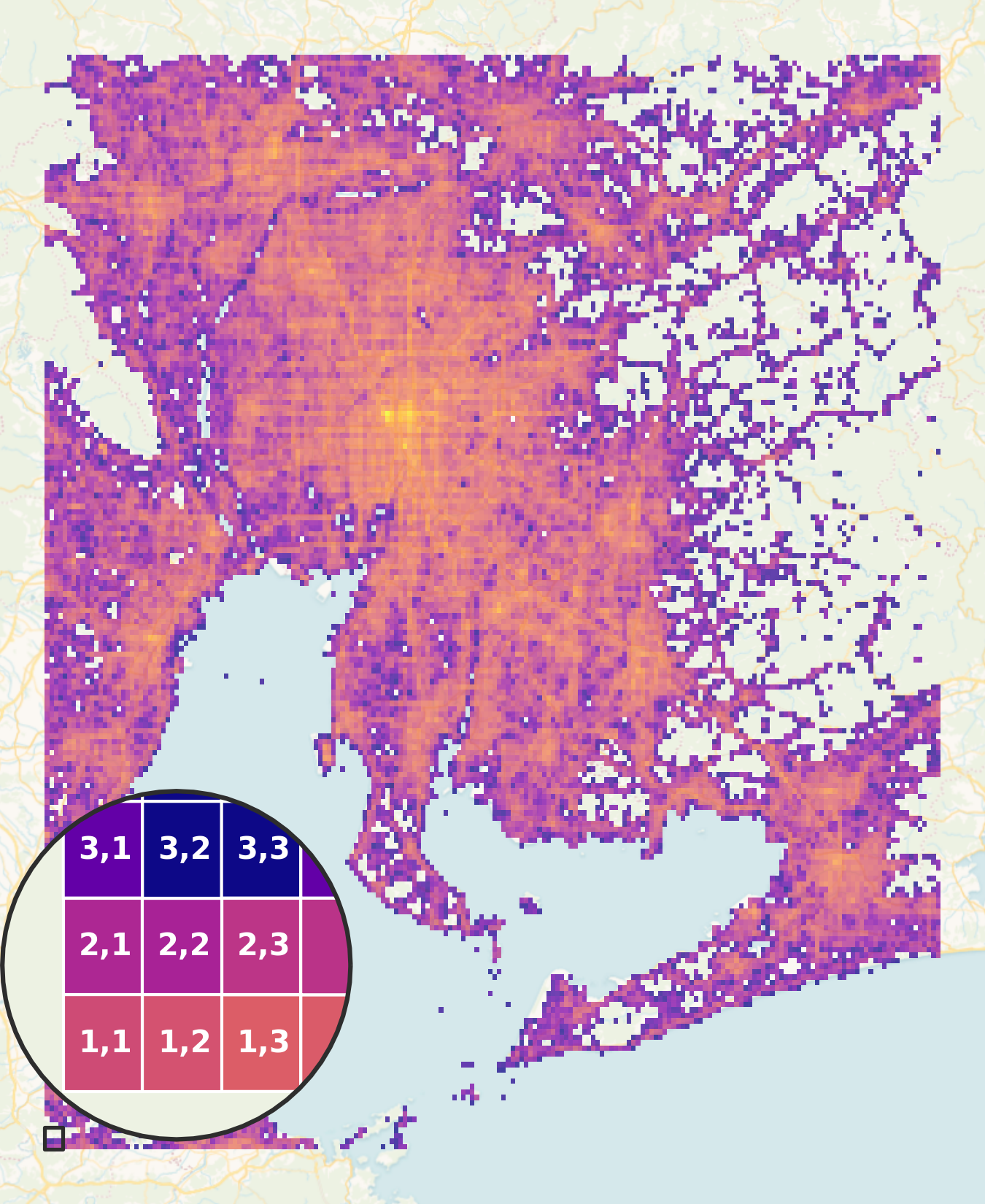}
        \captionsetup{position=bottom,justification=centering}
        \caption{}
        \label{fig:activity_count_log_on_map}
    \end{subfigure}
    \begin{subfigure}[t]{0.245\linewidth}
        \includegraphics[width=\linewidth]{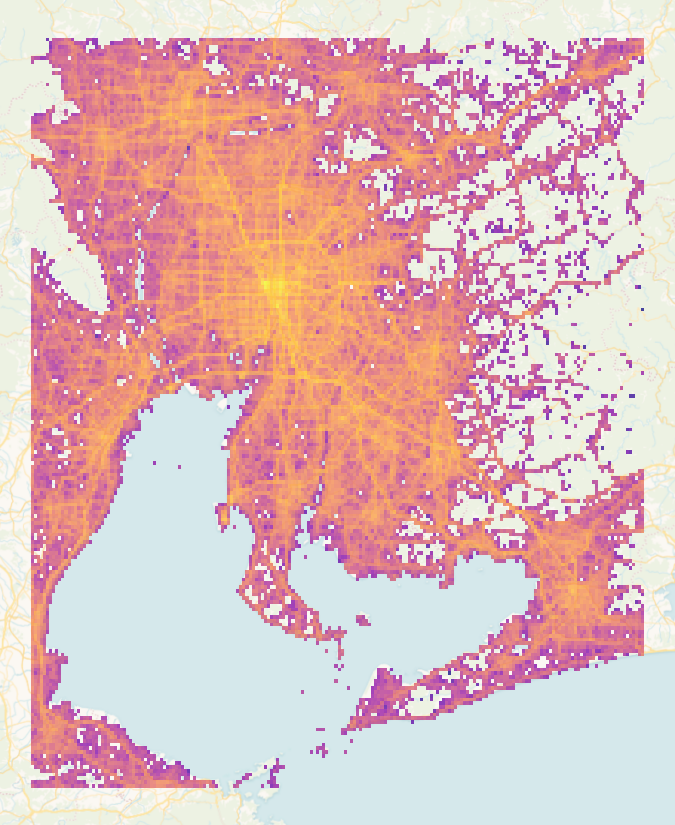}
        \captionsetup{position=bottom,justification=centering}
        \caption{}
        \label{fig:users_count_log_on_map}
    \end{subfigure}
    \begin{subfigure}[t]{0.245\linewidth}
        \includegraphics[width=\linewidth]{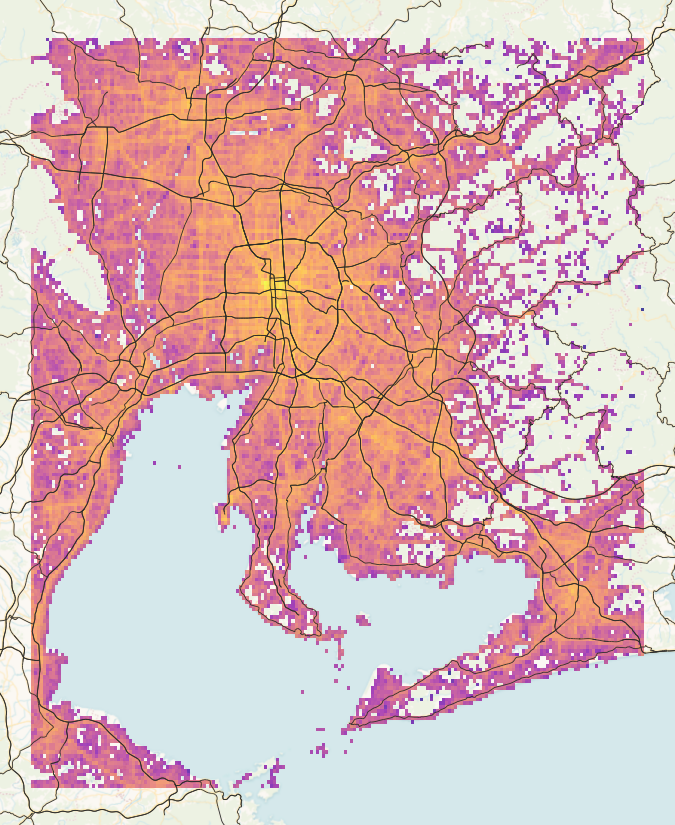}
        \captionsetup{position=bottom,justification=centering}
        \caption{}
        \label{fig:users_count_log_on_map_with_roads}
    \end{subfigure}
    \ifnum\includecomplexity = 1
    \begin{subfigure}[t]{0.245\linewidth}
        \includegraphics[width=\linewidth]{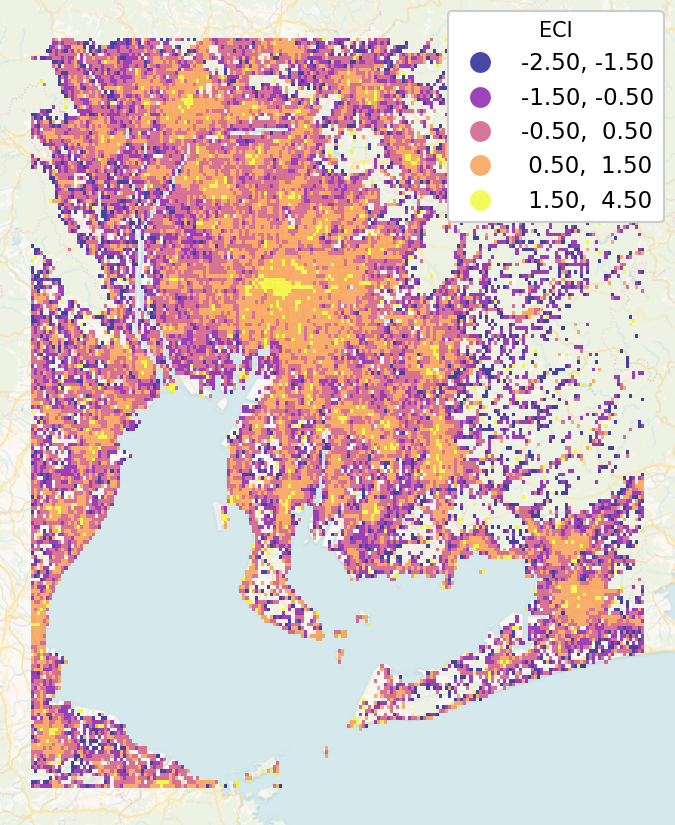}
        \captionsetup{position=bottom,justification=centering}
        \caption{}
        \label{fig:grid_complexity}
    \end{subfigure}
    \fi
    \caption{
        The reconstructed grid is plotted over a map, with colors indicating the number of activity records (\textbf{\subref{fig:activity_count_log_on_map}}) and unique users (\textbf{\subref{fig:users_count_log_on_map}}) on a log-scale, as well as the higher-order elements of the road network were also displayed (\textbf{\subref{fig:users_count_log_on_map_with_roads}})%
        \ifnum\includecomplexity = 1%
        , and the amenity complexity (\textbf{\subref{fig:grid_complexity}}) of the cells for additional details.
        \else%
        \,for additional details.
        \fi%
        To highlight the coastline, the cells were set to transparent if the activity value is below the same threshold as for Figure~\ref{fig:activity_cut}.
    }
    \label{fig:choropleths_on_map}
\end{figure}

The first step in the validation process is to plot the reconstructed grid over the map, with each cell colored according to the number of activity records (Figure~\ref{fig:activity_count_log_on_map}) and unique users (Figure~\ref{fig:users_count_log_on_map}).
The number of unique users per cell more accurately reflects the road network, particularly the highest order of highways (motorway, trunk) from the \acrshort*{OSM}, which were displayed as a validation (Figure~\ref{fig:users_count_log_on_map_with_roads}).
As the grid aligns with the map, it can be concluded that the grid geometry is considered good.

\ifnum\includecomplexity = 1
The amenity complexity, as defined by Juhász et al. \cite{juhasz2023amenity}, can be calculated using the provided the information about the number of different \acrshort*{POI}s (\acrlong*{POI}) in each cell.
Figure~\ref{fig:grid_complexity} shows the \acrfull*{ECI}, calculated using the `ecomplexity' Python package.
The highest \acrshort*{ECI} values correspond to the city center of Nagoya and the center of other cities.
\fi

Another direction of the grid validation could be the home detection.
It is common practice \cite{pappalardo2021evaluation,pinter2022awakening,vanhoof2018assessing} to apply home detection algorithms to mobility data.
In this case, the cell with the most activity after 21:00 and before 8:00 is considered the home location.
It should be noted that the home detection does not require the reconstructed grid it can be computed from the original data.
The thresholds (21:00 and 8:00) are set as a rule of thumb and may require adjustment to align with Japanese societal norms.
Furthermore, the home detection algorithm does not differentiate between workdays and weekends as the data does not contain dates.
However, holidays could be inferred from the daily activity levels.

\begin{figure}[t]
    \centering
    \begin{subfigure}[t]{0.40\linewidth}
        \includegraphics[width=\linewidth]{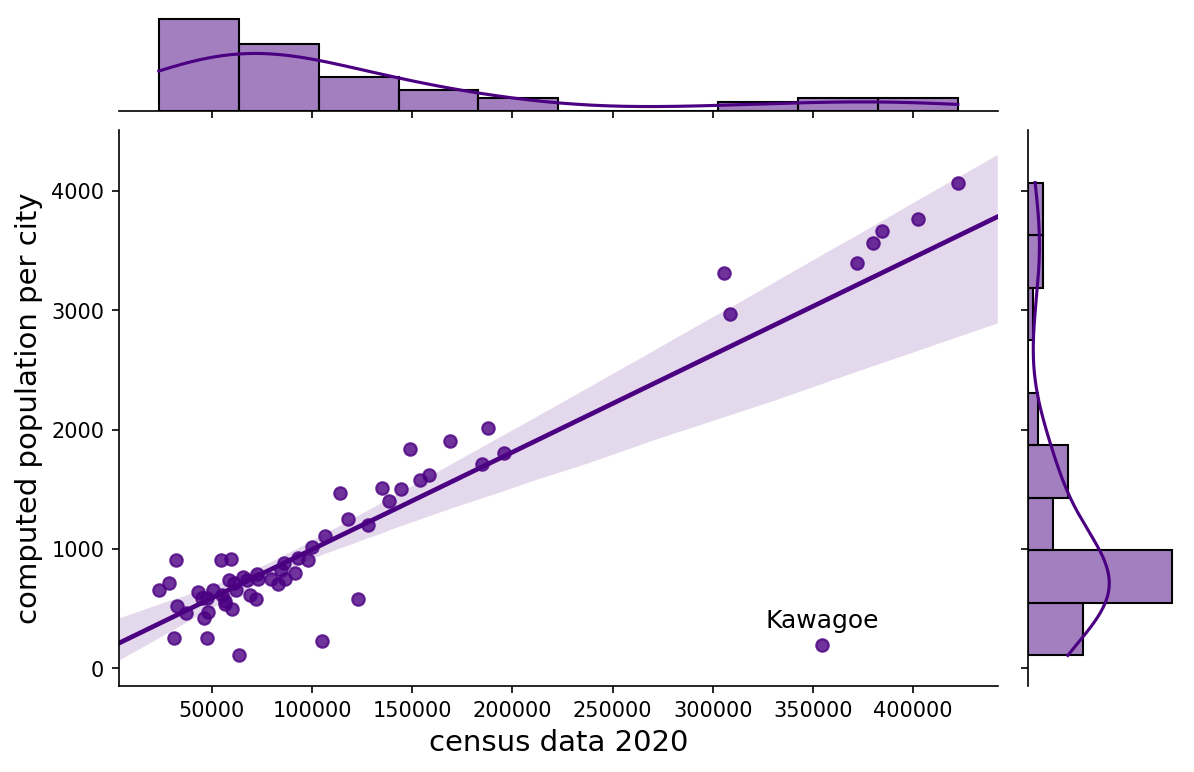}
        \captionsetup{position=bottom,justification=centering}
        \caption{}
        \label{fig:population_per_city}
    \end{subfigure}
    \begin{subfigure}[t]{0.40\linewidth}
        \includegraphics[width=\linewidth]{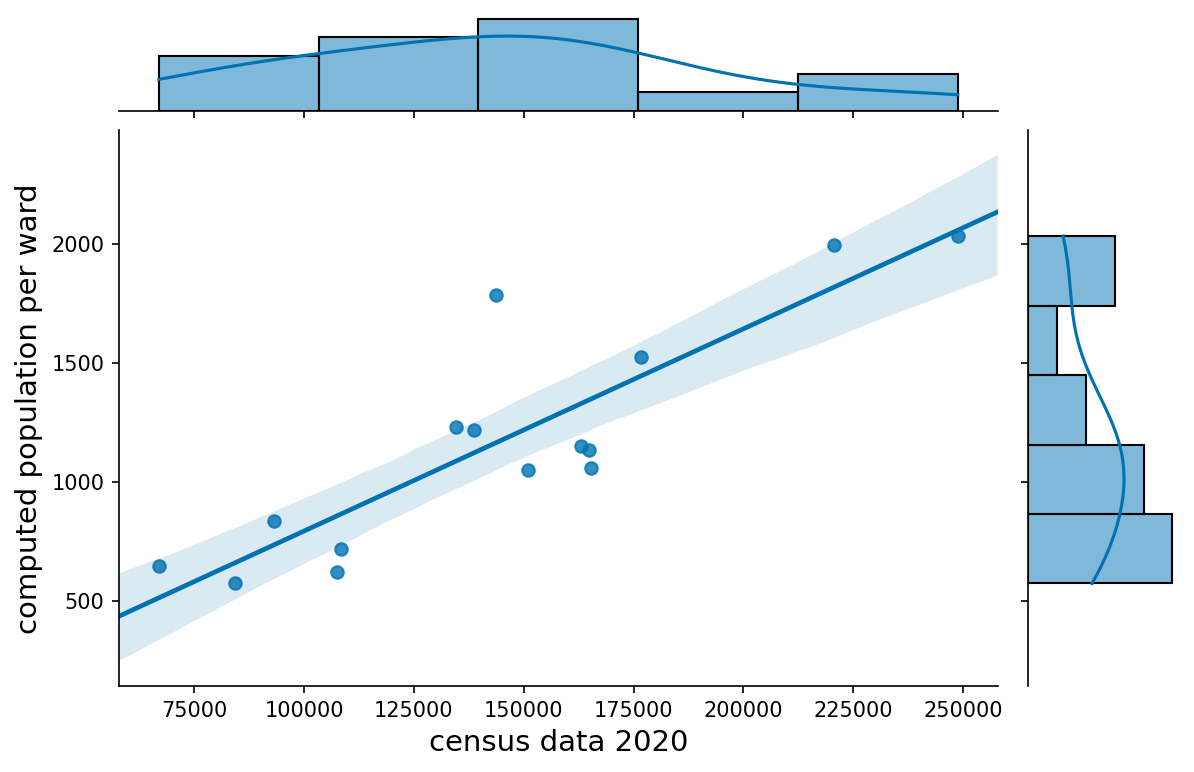}
        \captionsetup{position=bottom,justification=centering}
        \caption{}
        \label{fig:population_per_ward}
    \end{subfigure}
    \caption{
        The correlation between the population based on census data and the estimated number of inhabitants.
        The comparison is made at the municipal level (\textbf{\subref{fig:population_per_city}}) and also on the ward level of Nagoya (\textbf{\subref{fig:population_per_ward}}).
        Cities whose area is covered by the observation area with less than 30\% are excluded from the comparison.
    }
    \label{fig:population}
\end{figure}

Without ground truth, the detected home locations cannot be validated, but their spatial distribution can be compared \cite{juhasz2023amenity,pinter2022commuting} to the census data \cite{estat2020population}.
Figure~\ref{fig:population_per_city} compares the estimated number of inhabitants for the cities within the observation area with the 2020 census data.
The Pearson's R correlation coefficient is \num{\CityCorrelation}.
It should be noted that cities with less than 30\% of their area within the observation area have been excluded from the comparison.
Interestingly, Kawagoe is still an outlier (denoted in Figure~\ref{fig:population_per_city}) and appears to be underrepresented in the mobility data.
Upon removing Kawagoe from the comparison, the city-level correlation coefficient (Pearson’s R) increases to \num{\CityCorrelationWithoutKawagoe}.
The results show that the data follows approximately 1\% of the population.
Figure~\ref{fig:population_per_ward} shows the correlation at the ward level in Nagoya (Pearson's R is \num{\NagoyaWardCorrelation}).

The spatial distribution of the detected population correlates with the census data, thereby confirming the validity of the constructed grid geometry.
%
In contrast, in the case of the unstretched grid (Figure~\ref{fig:template_matched_zoomed_unstretched}), the correlations are notably worse.
Pearson's R is \num{\CityCorrelationUnstretched} for the municipality level and \num{\NagoyaWardCorrelationUnstretched} for the ward level.

\subsection*{Robustness}

\begin{figure}[th]
    \centering
    \begin{subfigure}[t]{0.32\linewidth}
        \includegraphics[width=\linewidth]{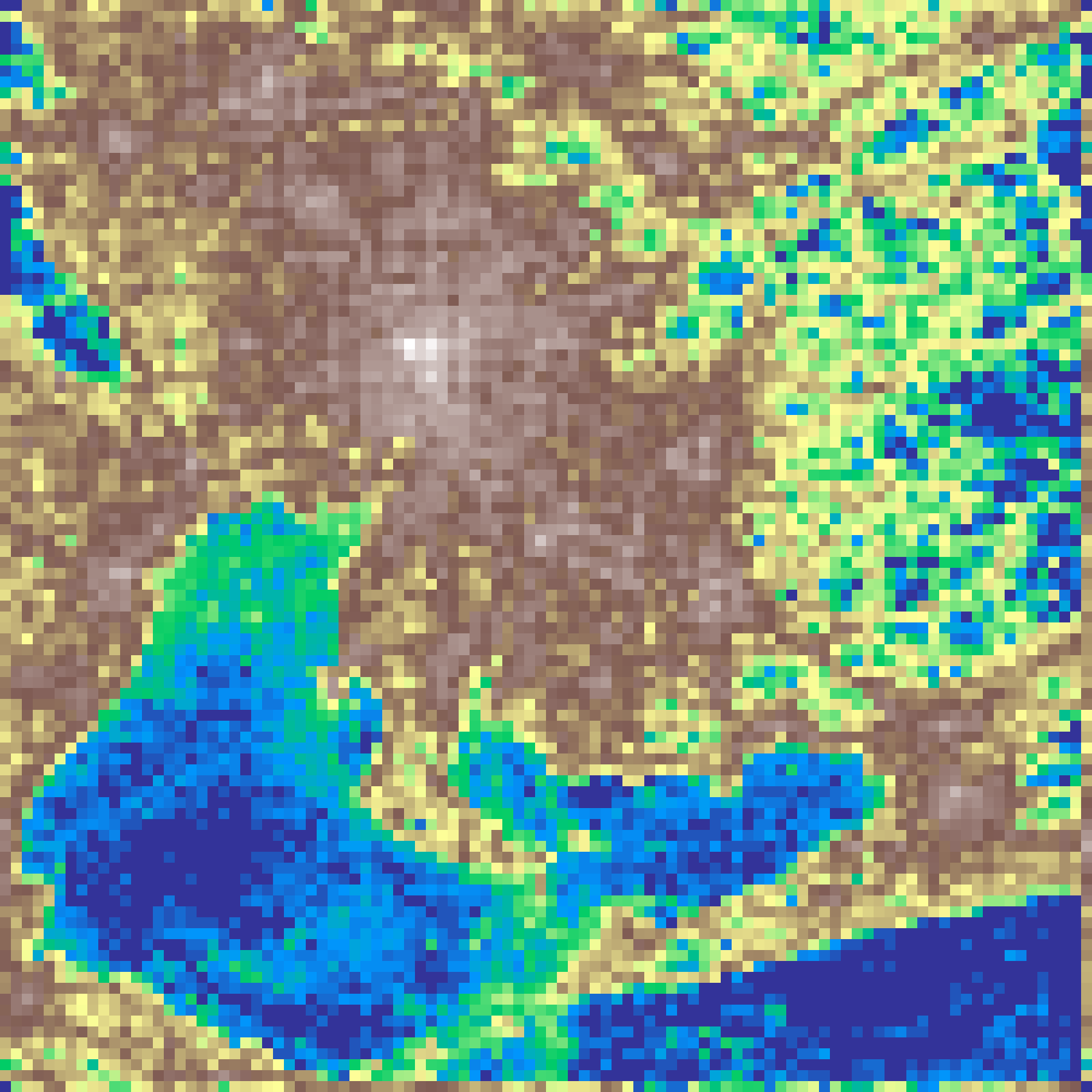}
        \captionsetup{position=bottom,justification=centering}
        \caption{}
        \label{fig:rescaled_2}
    \end{subfigure}
    \begin{subfigure}[t]{0.32\linewidth}
        \includegraphics[width=\linewidth]{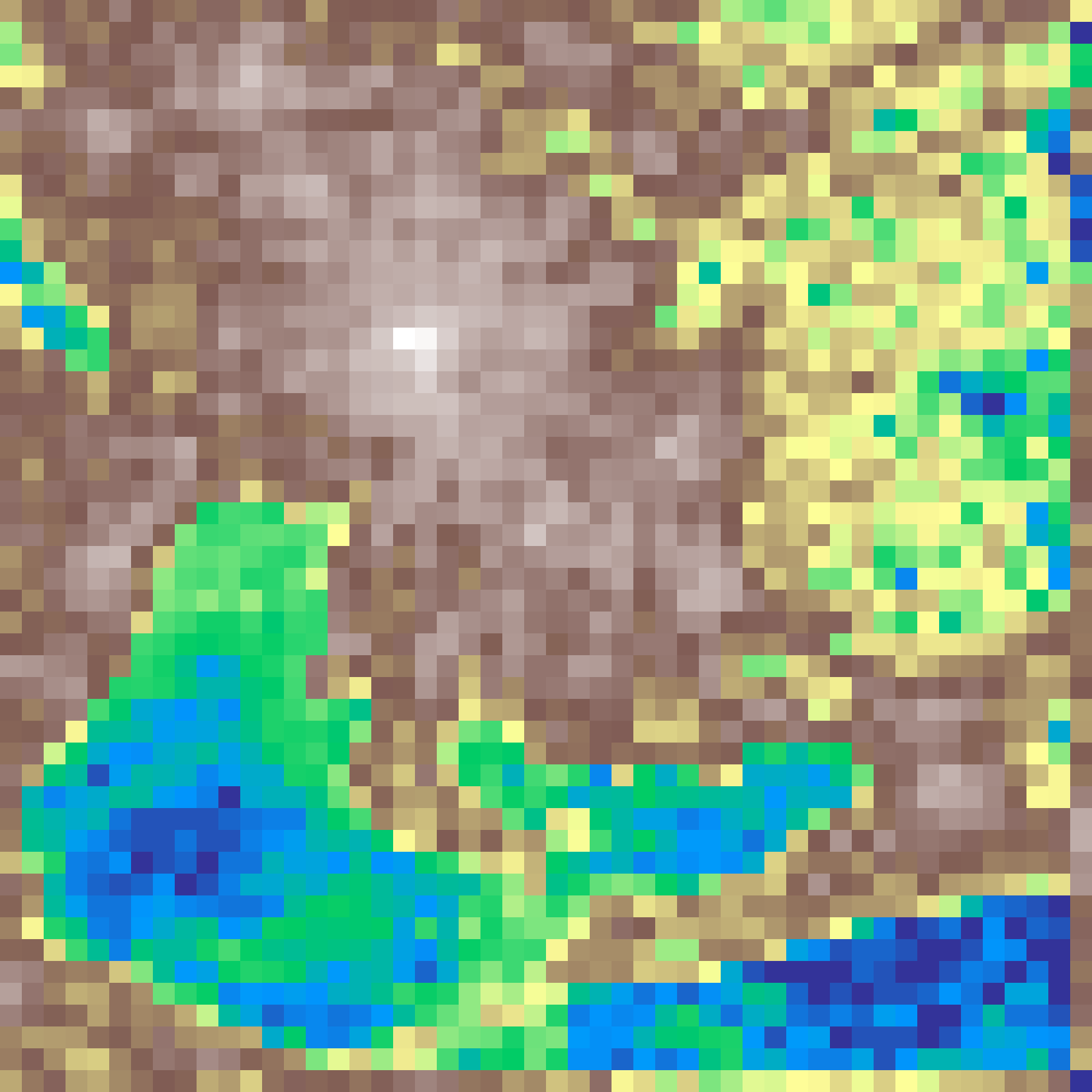}
        \captionsetup{position=bottom,justification=centering}
        \caption{}
        \label{fig:rescaled_4}
    \end{subfigure}
    \begin{subfigure}[t]{0.32\linewidth}
        \includegraphics[width=\linewidth]{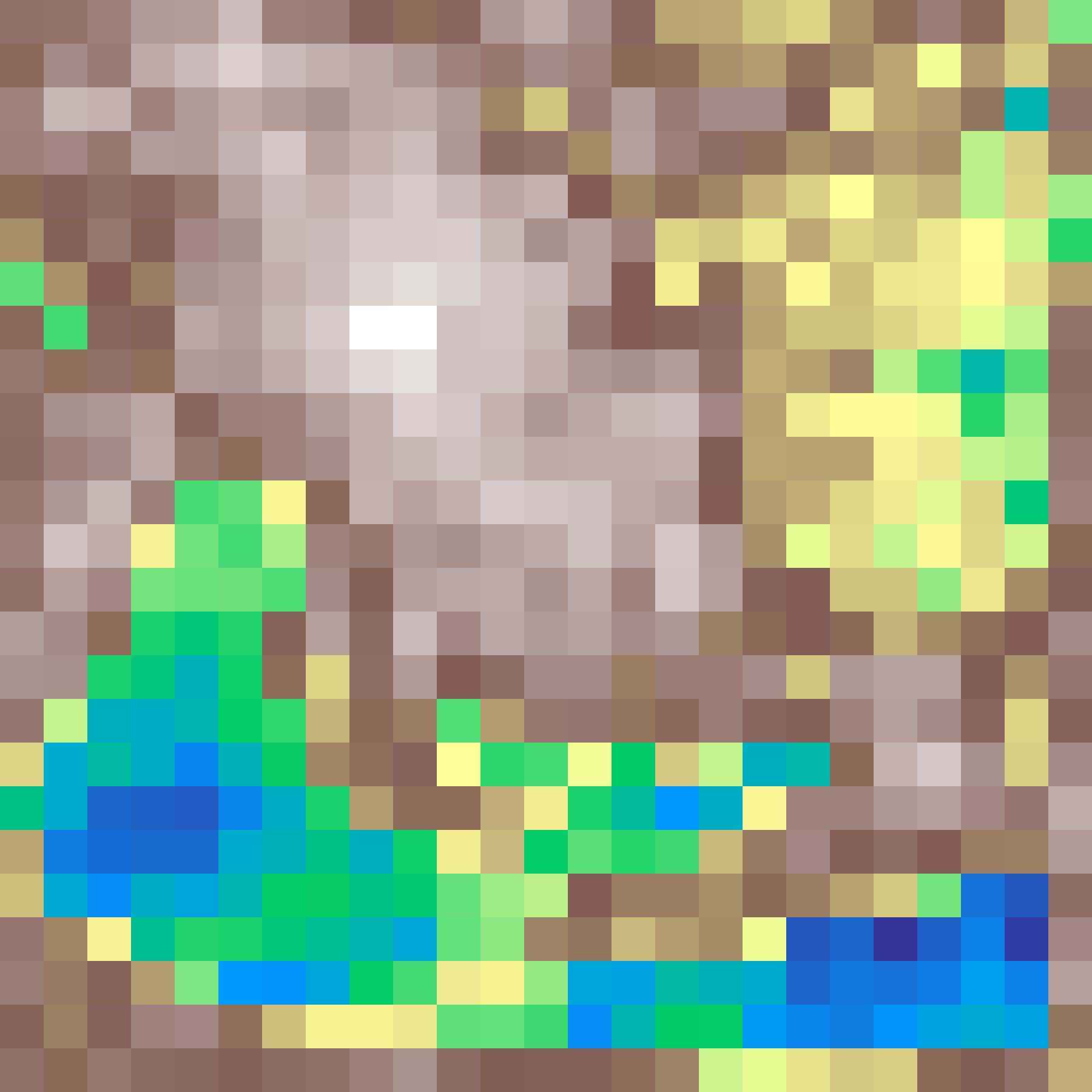}
        \captionsetup{position=bottom,justification=centering}
        \caption{}
        \label{fig:rescaled_8}
    \end{subfigure}
    \caption{
        The rescaled grids plotted as heatmaps. Instead of the original 500 m by 500 m grid, a 1~km by 1~km (\textbf{\subref{fig:rescaled_2}}), a 2~km by 2~km (\textbf{\subref{fig:rescaled_4}}), and a 4~km by~4 km (\textbf{\subref{fig:rescaled_8}}) cells are used, resulting $100\times100$, $50\times50$ and $25\times25$ element matrices.
    }
    \label{fig:scaling}
\end{figure}

It was previously demonstrated that the urban landscape can be inferred from mobility data even if the observation area is undisclosed.
To what extent is the presented method robust to increases in the resolution of the discretization?
The reconstructed grid was merged into grids of 1 km by 1 km, 2 km by 2 km, and 4 km by 4 km grids by summing the activity records in 4, 16, and 64 cells, respectively.
An increase in the cell size results in a reduction in the accuracy of user location, thereby increasing privacy.
Figure~\ref{fig:scaling} shows the rescaled grids as heatmaps using the same color scheme as in Figure~\ref{fig:user_heatmap_terrain_fixed}.
Although the rescaled grids notably compress the information and the figures lose details, the urban landscape remains more or less recognizable.
The expectation is that the presented approach, utilizing template matching, will be able of accurately identifying the urban area.

\begin{figure}[th]
    \centering
    \begin{subfigure}[t]{0.245\linewidth}
        \includegraphics[width=\linewidth, height=4cm]{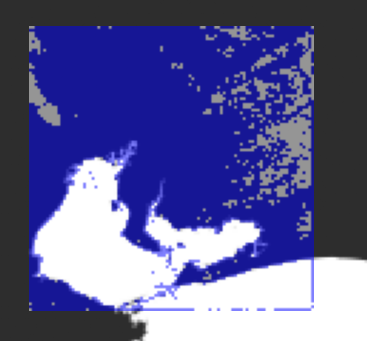}
        \captionsetup{position=bottom,justification=centering}
        \caption{}
        \label{fig:rescaled_result_1k}
    \end{subfigure}
    \begin{subfigure}[t]{0.245\linewidth}
        \includegraphics[width=\linewidth, height=4cm]{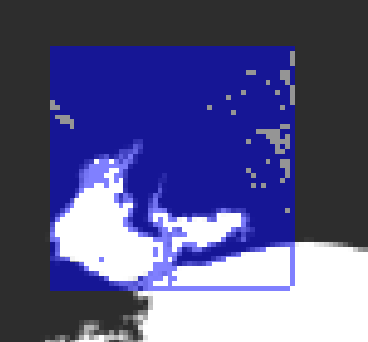}
        \captionsetup{position=bottom,justification=centering}
        \caption{}
        \label{fig:rescaled_result_2k}
    \end{subfigure}
    \begin{subfigure}[t]{0.245\linewidth}
        \includegraphics[width=\linewidth, height=4cm]{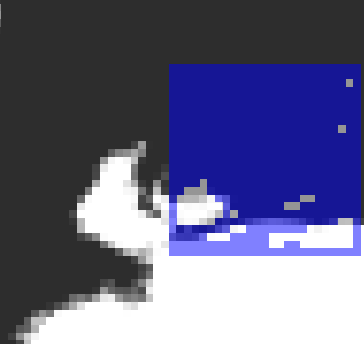}
        \captionsetup{position=bottom,justification=centering}
        \caption{}
        \label{fig:rescaled_result_4k}
    \end{subfigure}
    \begin{subfigure}[t]{0.245\linewidth}
        \includegraphics[width=\linewidth, height=4cm]{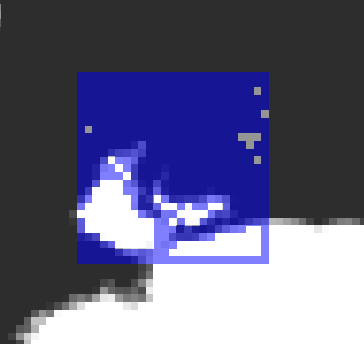}
        \captionsetup{position=bottom,justification=centering}
        \caption{}
        \label{fig:rescaled_result_4kv2}
    \end{subfigure}
    \caption{
        The results of the  template matching for the rescaled grids with the original template threshold (75) 1~km by 1~km (\textbf{\subref{fig:rescaled_result_1k}}), 2~km by 2~km (\textbf{\subref{fig:rescaled_result_2k}}), and 4~km by~4 km (\textbf{\subref{fig:rescaled_result_4k}}).
        Additionally, the 4~km by~4 km result with an increased threshold (\textbf{\subref{fig:rescaled_result_4kv2}}).
    }
    \label{fig:scaled_results}
\end{figure}

The template was generated with the same threshold (75) as for the original grid (Figure~\ref{fig:thresholding}).
Figure~\ref{fig:scaled_results} shows the location results.
In the first two cases (Figure~\ref{fig:rescaled_result_1k} and \ref{fig:rescaled_result_2k}), the locating was accurate.
However, it failed for the 4 km by 4 km grid (Figure~\ref{fig:rescaled_result_4k}).
The threshold proved too low concerning the compressed image as the shape of Mikawa and Ise bays was lost, so the threshold was increased to 375.
Upon increasing the threshold, the template matching algorithm was able to successfully locate the urban area (Figure~\ref{fig:rescaled_result_4kv2}).
This result proves that the characteristics of human mobility can reveal the urban landscape even when the mobility locations are discretized into 4 km by 4 km cells.

Another direction of the robustness checks is extending the presented method to other cities.
The openly available Weeplaces data set \cite{chen2022weeplaces} was used, presented in \cite{chen2022contrasting}, which was collected from \acrfull*{LBSN} including Facebook Places, Foursquare, and Gowalla.
Two cities were selected from the Weeplaces database for further analysis: Toronto and London.
The activity locations were discretized into a 500 m by 500 m grid over a 100 km by 100 km by area, which also covers other settlements from the area similarly as in the YJMob100K data set \cite{yabe2023dataset}.
Note that an area of this size covers other settlements outside Toronto from the shore of Lake Ontario.
The Weeplaces data set contains a significantly lower number of activity records per city than the YJMob100K data set.
In the case of Toronto, there are \num{\TorontoActivityCount} activity records from \num{\TorontoNumberOfUniqueUsers} unique users in the selected area, whereas the YJMob100K data set contains \num{\NagoyaActivityCount} activity records from \num{100000} users.

\begin{figure}[th]
    \centering
    \begin{subfigure}[t]{0.225\linewidth}
        \includegraphics[width=\linewidth]{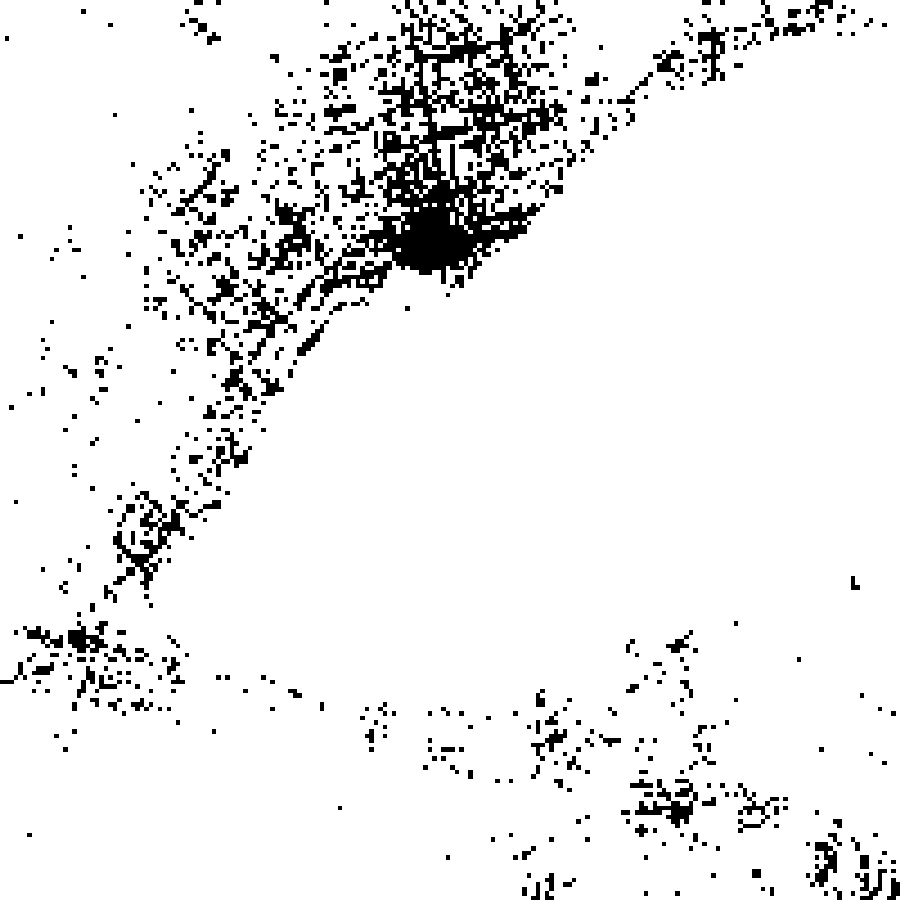}
        \captionsetup{position=bottom,justification=centering}
        \caption{}
        \label{fig:toronto_grid}
    \end{subfigure}
    \begin{subfigure}[t]{0.225\linewidth}
        \includegraphics[width=\linewidth]{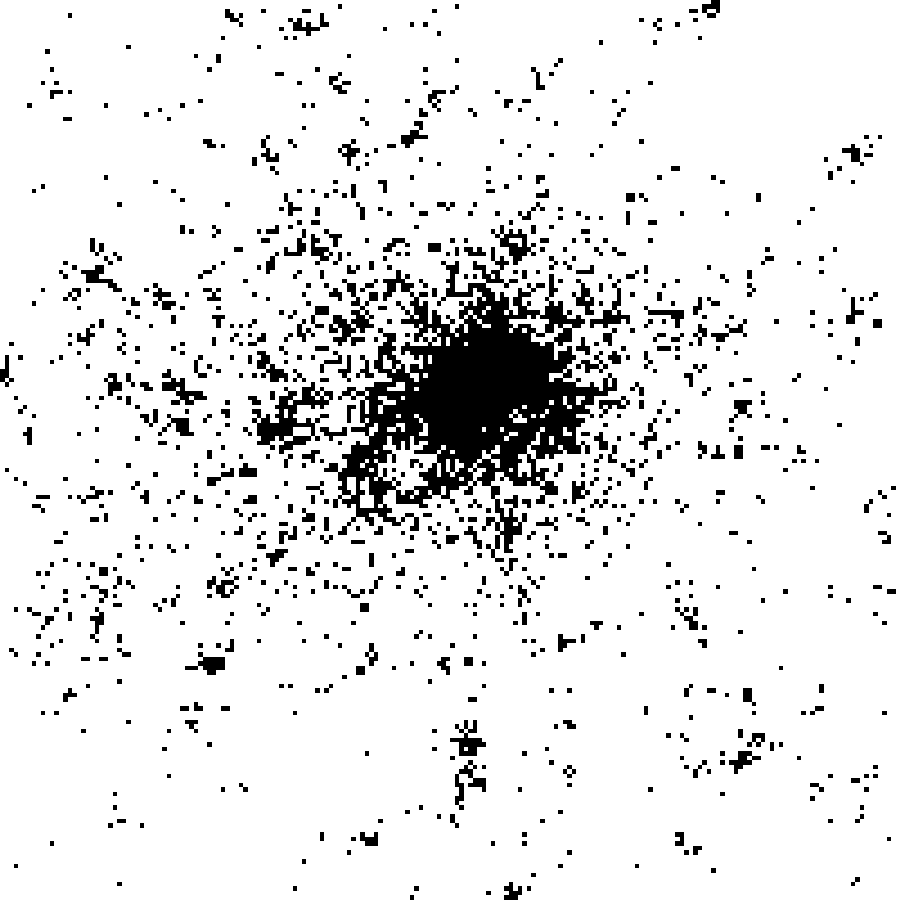}
        \captionsetup{position=bottom,justification=centering}
        \caption{}
        \label{fig:london_grid}
    \end{subfigure}
    \begin{subfigure}[t]{0.225\linewidth}
        \includegraphics[width=\linewidth]{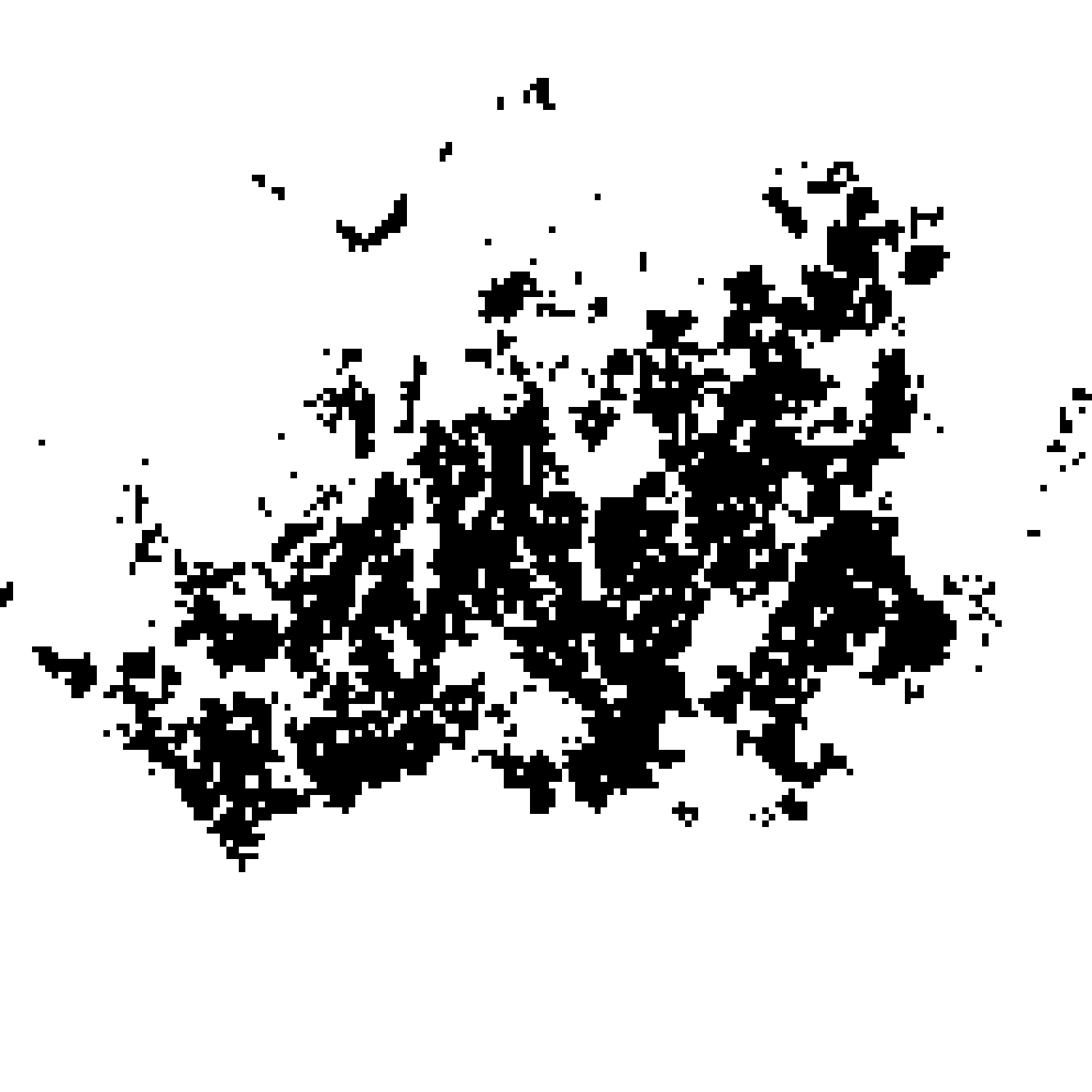}
        \captionsetup{position=bottom,justification=centering}
        \caption{}
        \label{fig:helsinki_grid}
    \end{subfigure}
    \begin{subfigure}[t]{0.225\linewidth}
        \includegraphics[width=\linewidth]{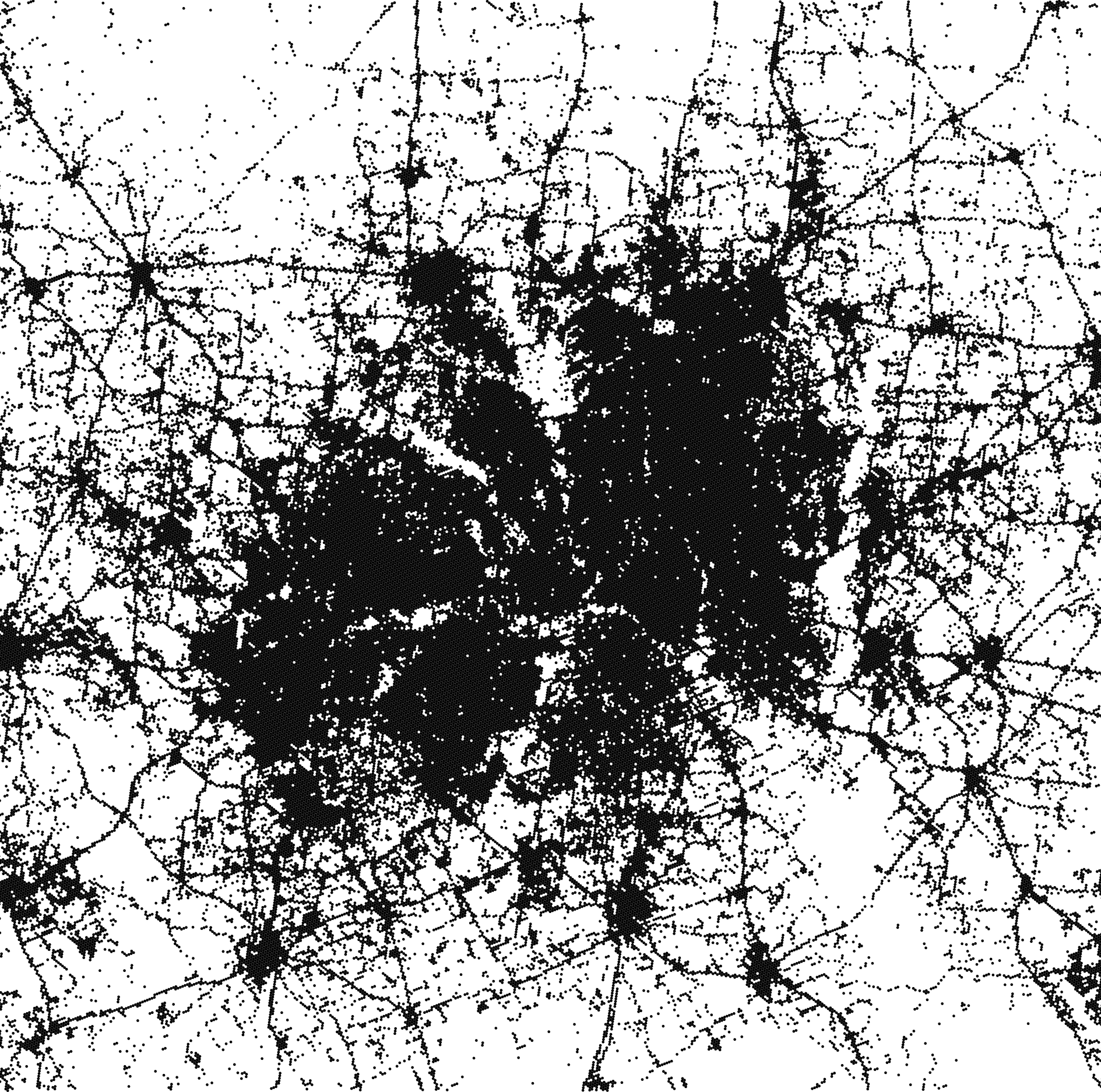}
        \captionsetup{position=bottom,justification=centering}
        \caption{}
        \label{fig:dallas_data}
    \end{subfigure}

    \begin{subfigure}[t]{0.225\linewidth}
        \includegraphics[width=\linewidth]{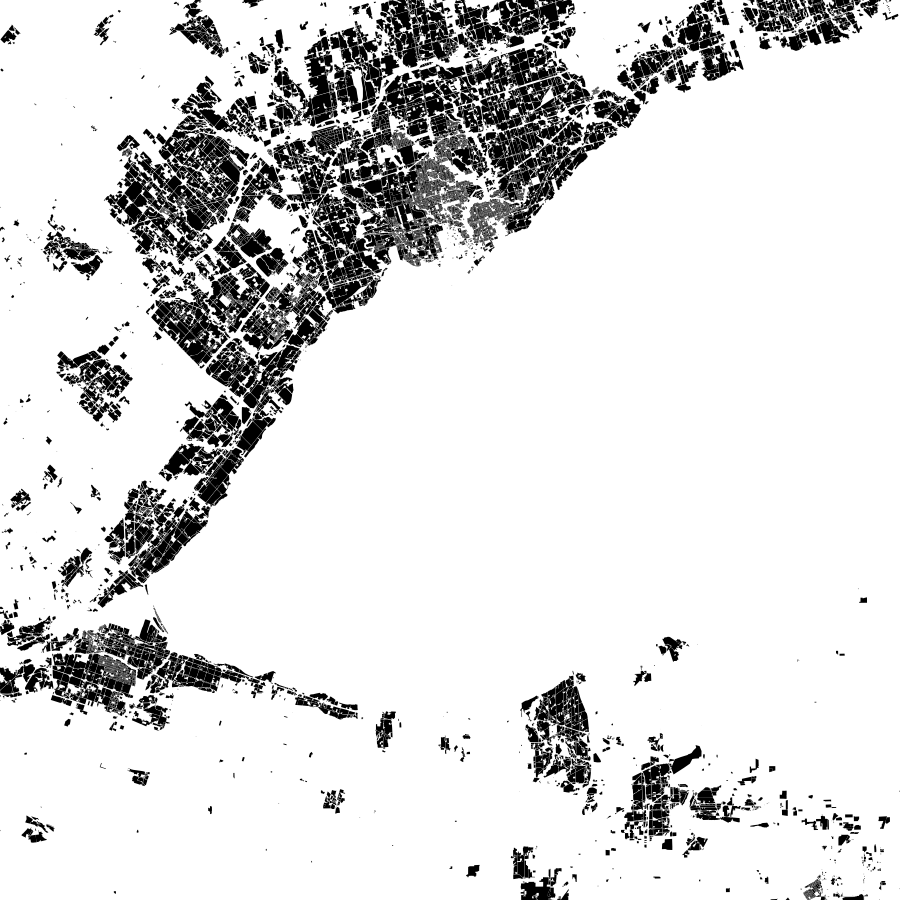}
        \captionsetup{position=bottom,justification=centering}
        \caption{}
        \label{fig:toronto_landuse}
    \end{subfigure}
    \begin{subfigure}[t]{0.225\linewidth}
        \includegraphics[width=\linewidth]{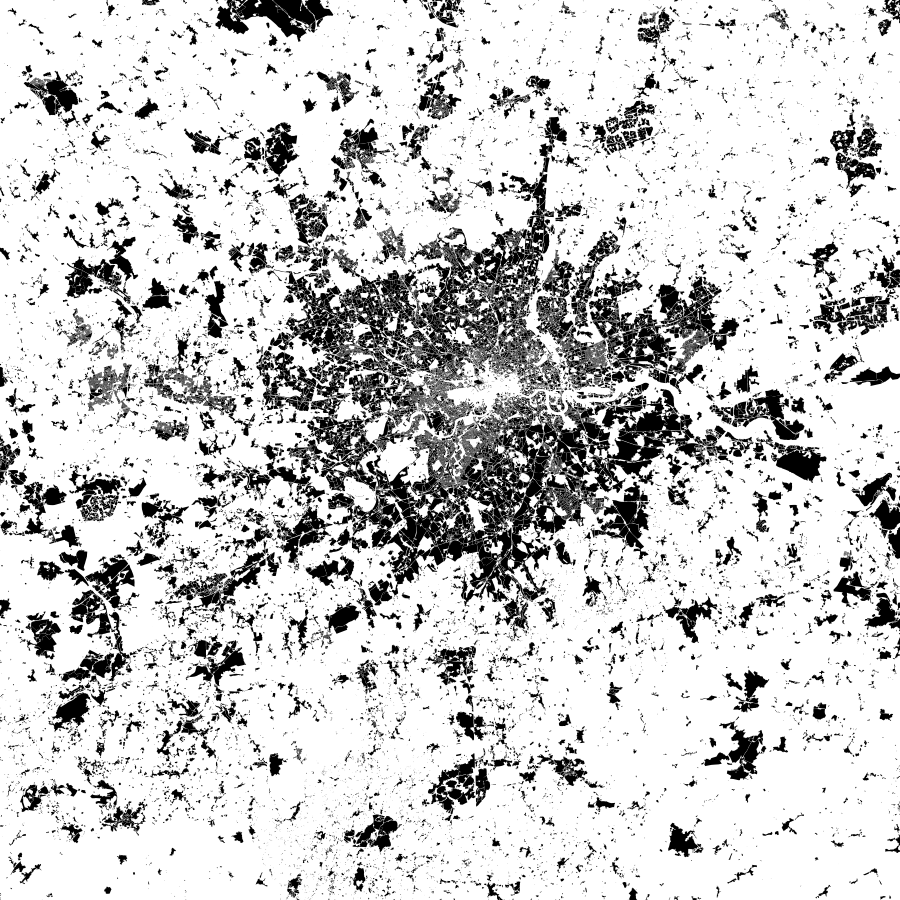}
        \captionsetup{position=bottom,justification=centering}
        \caption{}
        \label{fig:london_landuse}
    \end{subfigure}
    \begin{subfigure}[t]{0.225\linewidth}
        \includegraphics[width=\linewidth]{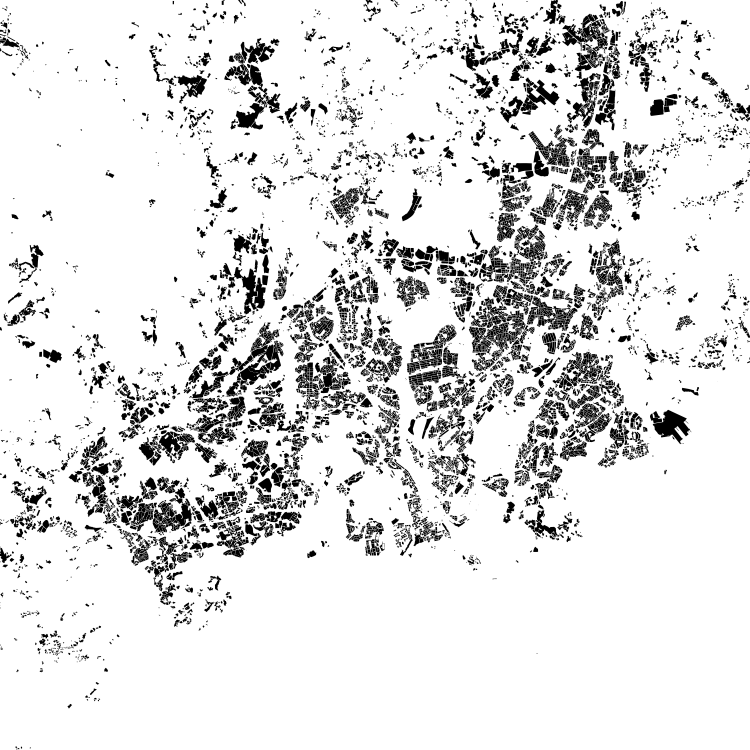}
        \captionsetup{position=bottom,justification=centering}
        \caption{}
        \label{fig:helsinki_landuse}
    \end{subfigure}
    \begin{subfigure}[t]{0.225\linewidth}
        \includegraphics[width=\linewidth]{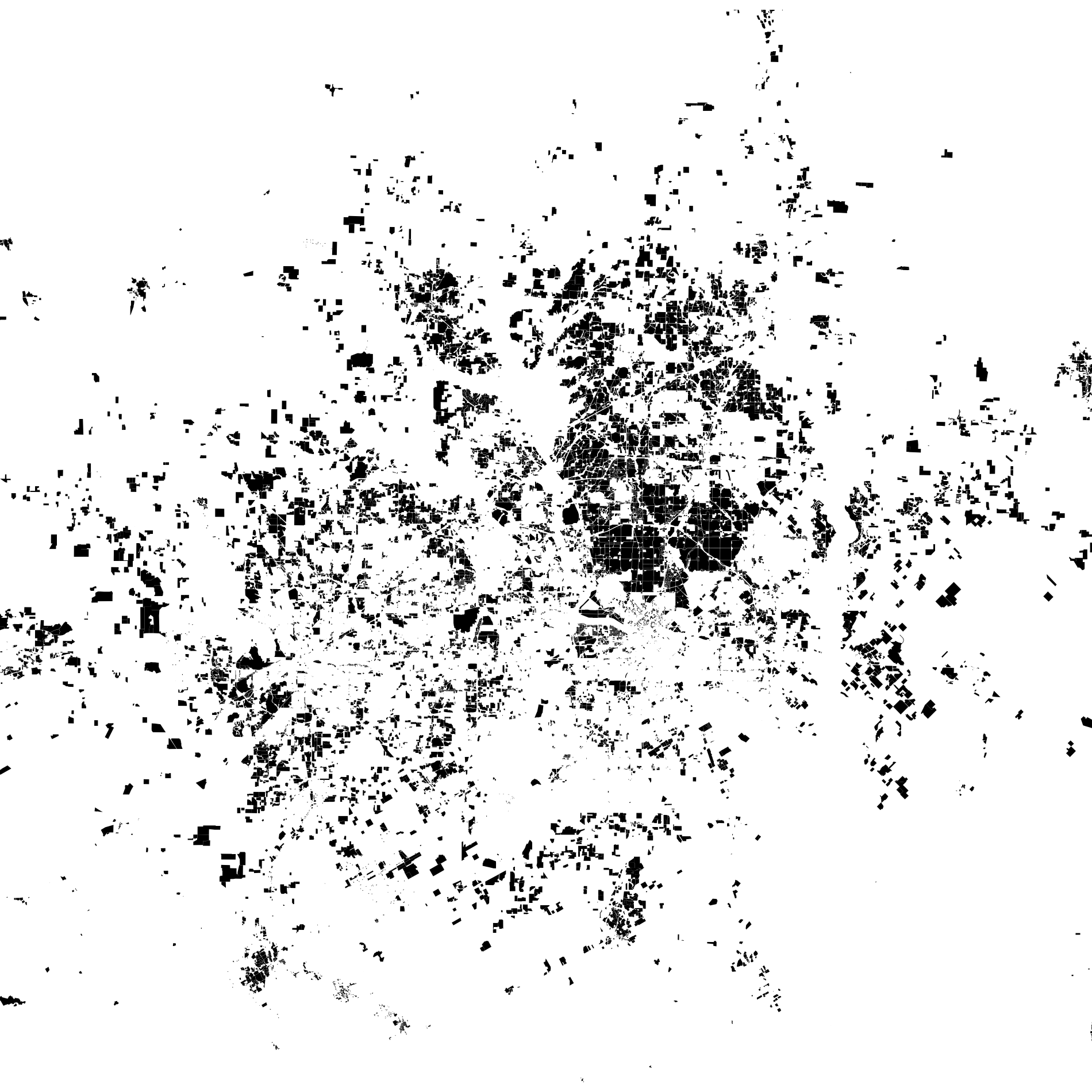}
        \captionsetup{position=bottom,justification=centering}
        \caption{}
        \label{fig:dallas_landuse}
    \end{subfigure}

    \begin{subfigure}[t]{0.225\linewidth}
        \includegraphics[width=\linewidth]{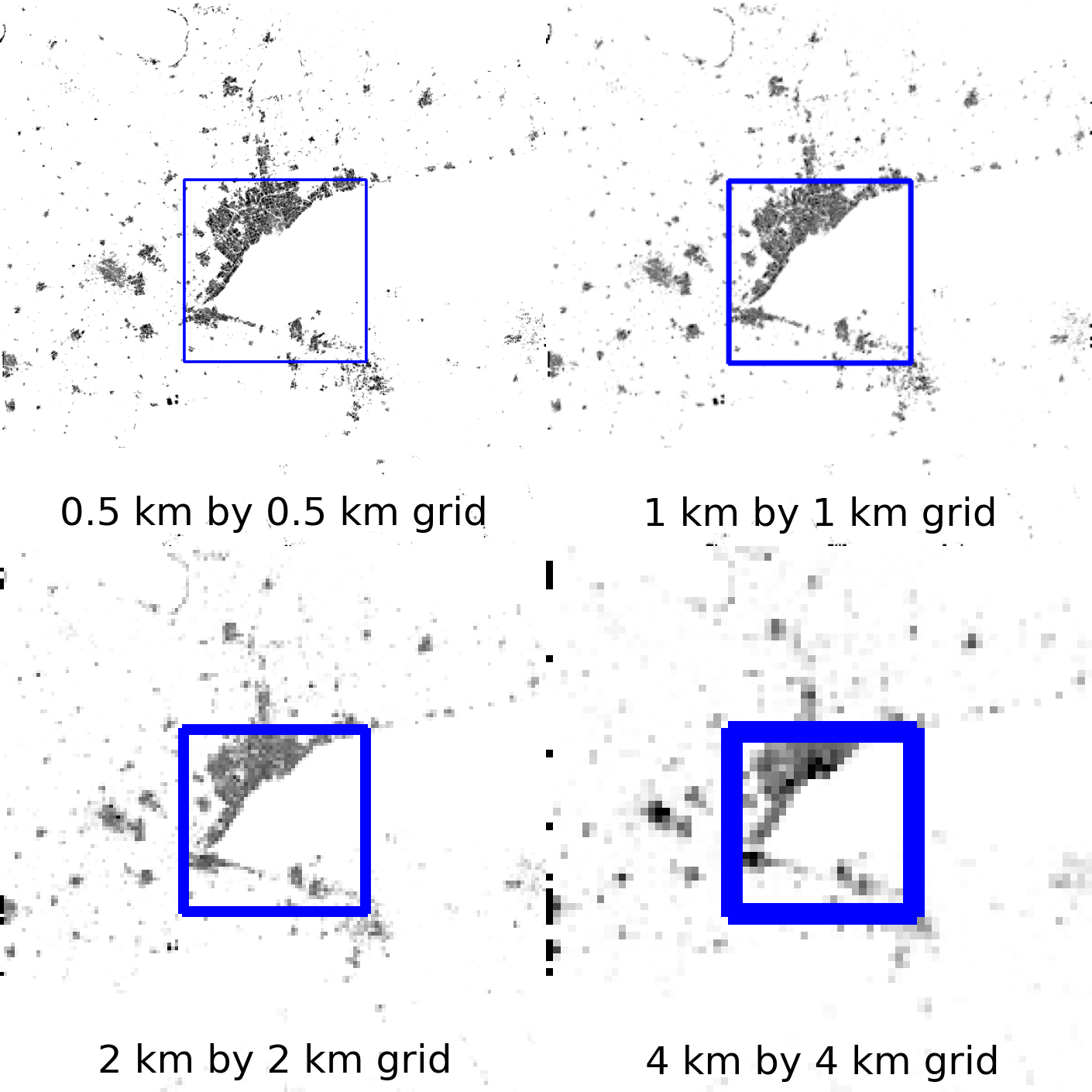}
        \captionsetup{position=bottom,justification=centering}
        \caption{}
        \label{fig:toronto_located}
    \end{subfigure}
    \begin{subfigure}[t]{0.225\linewidth}
        \includegraphics[width=\linewidth]{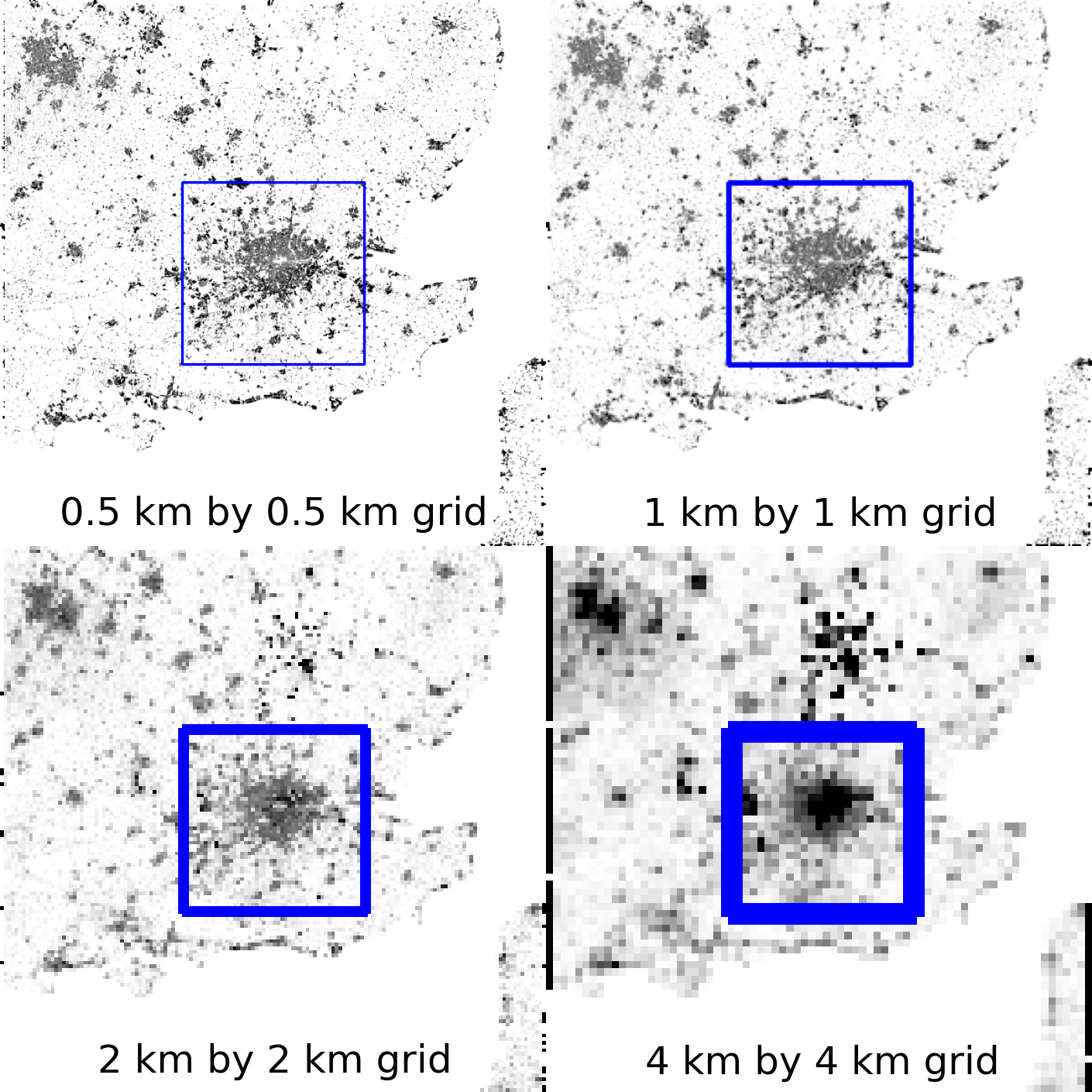}
        \captionsetup{position=bottom,justification=centering}
        \caption{}
        \label{fig:london_located}
    \end{subfigure}
    \begin{subfigure}[t]{0.225\linewidth}
        \includegraphics[width=\linewidth]{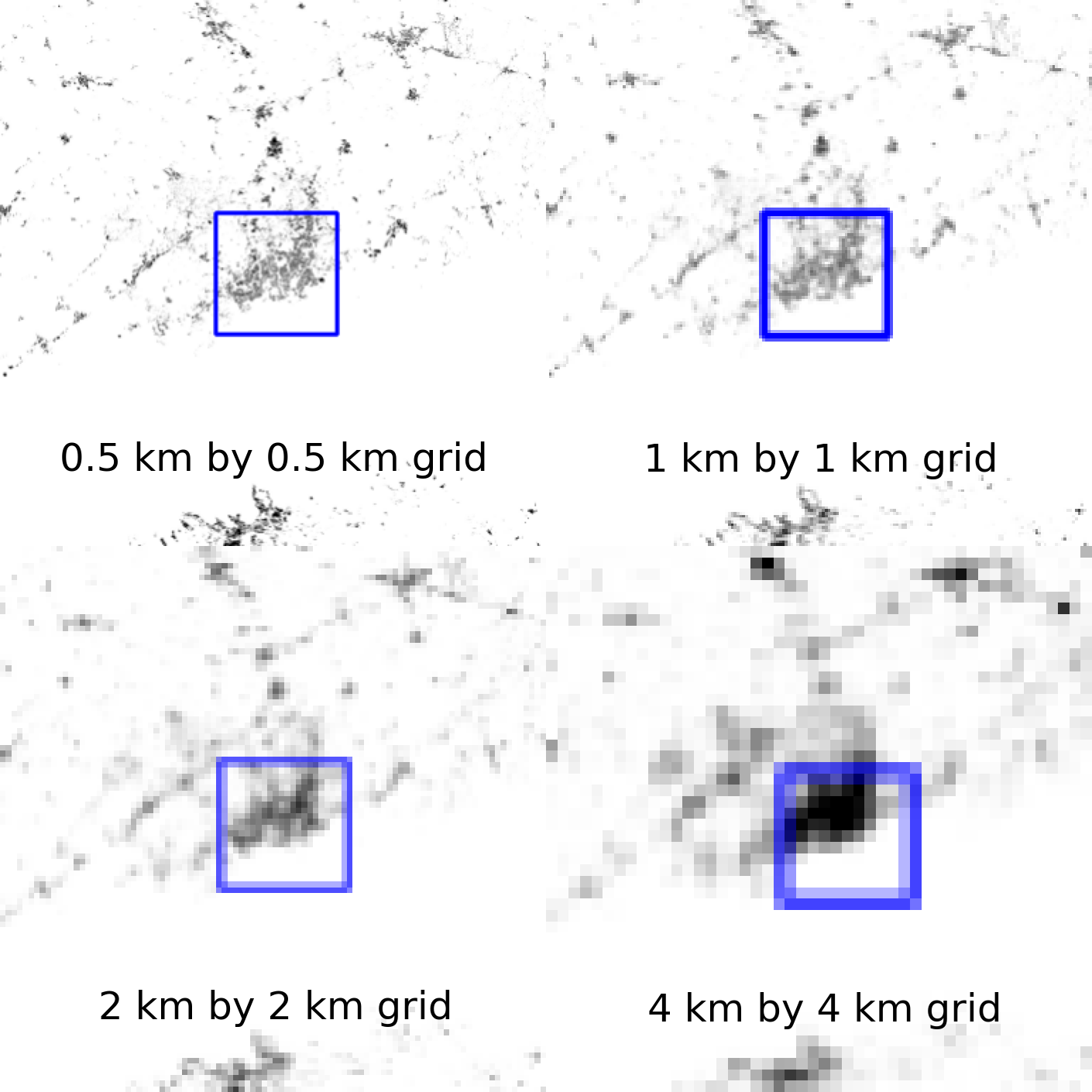}
        \captionsetup{position=bottom,justification=centering}
        \caption{}
        \label{fig:helsinki_located}
    \end{subfigure}
    \begin{subfigure}[t]{0.225\linewidth}
        \includegraphics[width=\linewidth]{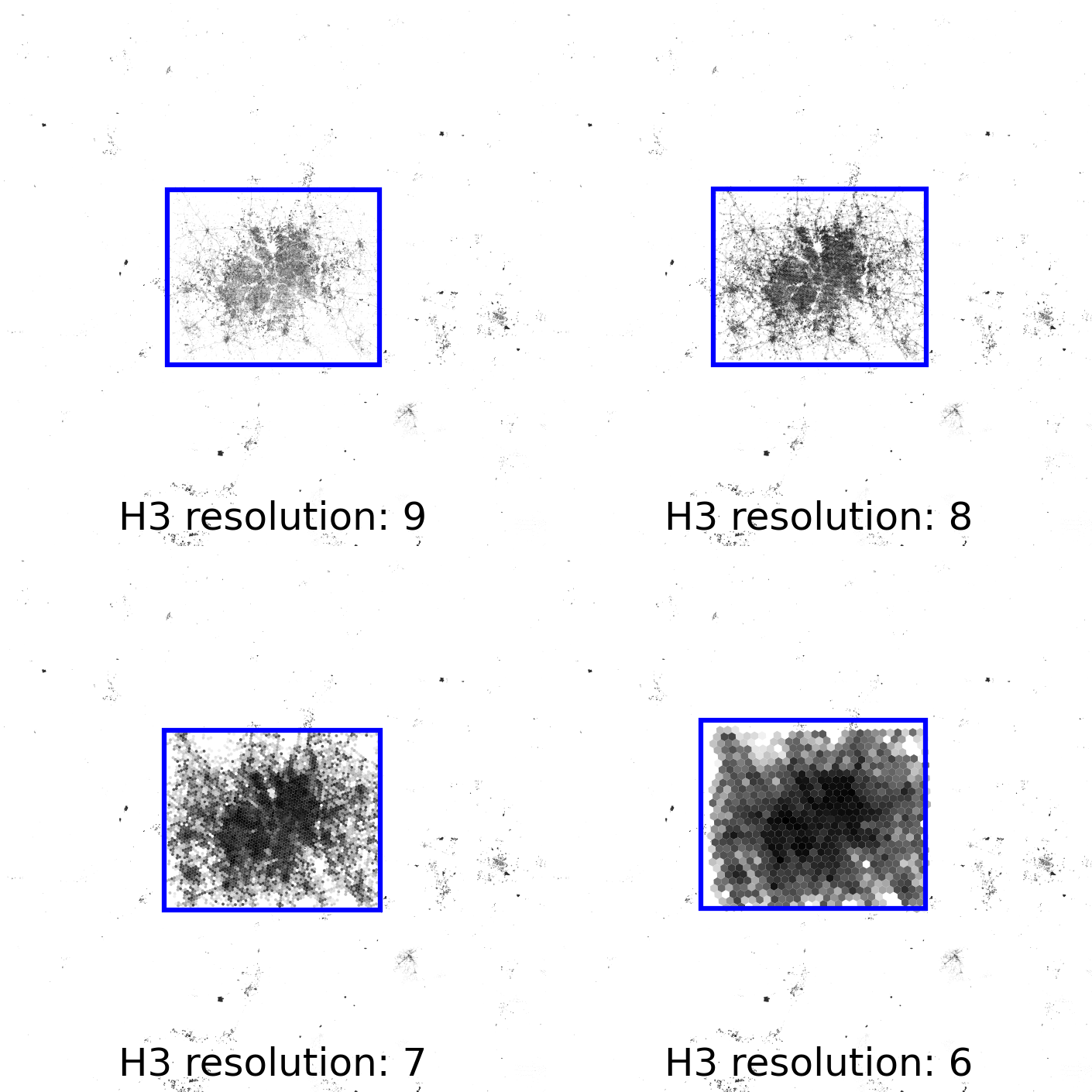}
        \captionsetup{position=bottom,justification=centering}
        \caption{}
        \label{fig:dallas_located}
    \end{subfigure}
    \caption{
        User activity from the Weeplaces data set discretized into a 500 m by 500 m grid around Toronto (\textbf{\subref{fig:toronto_grid}}), London (\textbf{\subref{fig:london_grid}}), Helsinki with a 250 m by 250 m grid from the population distribution data set (\textbf{\subref{fig:helsinki_grid}}), and pings from the Dallas--Fort Worth metroplex (\textbf{\subref{fig:dallas_data}}).
        The residential, retail, and industrial areas extracted from the \acrlong*{OSM} (\textbf{\subref{fig:toronto_landuse}}, \textbf{\subref{fig:london_landuse}}, \textbf{\subref{fig:helsinki_landuse}}, and \textbf{\subref{fig:dallas_landuse}}, respectively).
        The located urban areas of Toronto (\textbf{\subref{fig:toronto_located}}), London (\textbf{\subref{fig:london_located}}) and Helsinki (\textbf{\subref{fig:helsinki_located}}), using 500 m, 1 km, 2 km and 4 km squares, and  H3 hexagons at resolutions of 9, 8, 7 and 6 for the Dallas--Fort Worth metroplex (\textbf{\subref{fig:dallas_located}}) as the discretization method.
    }
    \label{fig:weeplaces}
\end{figure}

The Weeplaces data set is notable sparser, with significantly less activity concentrated outside the urban areas (Figure~\ref{fig:toronto_grid}), in contrast to the YJMob100K data set, where unpopulated areas still have some activity, even the ferry lines can be recognized.
The sparseness of the available data for Toronto makes it more difficult to compare the activity heatmap to the land.
Furthermore, cities without distinctive coastal regions, such as London (Figure~\ref{fig:london_grid}), may not work.
Alternatively, land usage data was extracted from \acrshort*{OSM}, namely regions designated as residential, retail, or industrial (Figure~\ref{fig:toronto_landuse} and \ref{fig:london_landuse}).
This approach aligns with the functional classification of the urban areas with respect to land cover data presented in \cite{bergroth202224}.
In this case, the template image contains black if the cell has any activity and white otherwise.
Figure~\ref{fig:toronto_located} and \ref{fig:london_located} illustrate that the urban area can be located using 500 m, 1 km, 2 km, and 4 km squares using the same method as in the case of the Nagoya metropolitan area (Figure~\ref{fig:scaled_results}).

A temporal density dataset was published \cite{bergroth202224} from the Helsinki Metropolitan Area.
It uses a 250 m by 250 m grid for the spatial dimension, which contains the portion of the population that was present in a given cell by hours.
For the template, the distribution values from the workday table at 21:00 were used, and the threshold value was \num{0.0025} (Figure\ref{fig:helsinki_grid}).
The template was compared to the land usage data from \acrshort*{OSM} (Figure\ref{fig:helsinki_landuse}).
Figure~\ref{fig:helsinki_located} shows the result of the template matching using 500 m, 1 km, 2 km, and 4 km grids.

Another data set was used to demonstrate the city landscape recognizability, covering the Dallas--Fort Worth metroplex \cite{makris2023dfw} from the paper \cite{makris2023mechanical}, which contains hourly locations of people.
As opposed to the previous procedure, the coordinates are discretized using H3 hexagons \cite{h3}, and the activity aggregated to different sizes of H3 hexagons are compared to the land usage (Figure~\ref{fig:dallas_landuse}).
The H3 resolutions between 6 and 9 are tested.
Larger resolutions represent smaller hexagons: a resolution-9 hexagon is about \num{0.1053} km\textsuperscript{2}, a resolution-8 hexagon is about \num{0.7373} km\textsuperscript{2}, a resolution-7 hexagon is about \num{5.1613} km\textsuperscript{2}, and a resolution-6 hexagon is about  \num{36.129} km\textsuperscript{2}.
The template matching algorithm determined the city location well in the three higher resolution cases, but slightly misplaced the template when using the resolution-6 hexagons  (Figure~\ref{fig:dallas_located}).

\subsection*{User identifiability}

In addition to the location identifiability, it is worth testing the user identifiability using different grid scales.
A user is considered identified by their top four locations, similar to \cite{zang2011anonymization} where location granularity (cell, zip code, city, county, and state) was tested by the top 1, 2, and 3 locations.
In this paper, 1 km, 2 km, 4 km, 8 km, and 16 km square grids were used.
The activities were remapped to the upscaled grids, and the top four locations were determined.
A user is considered identifiable in the upscaled grid if the most visited four locations can still be distinguished.
Table~\ref{tab:top4} shows the number of identifiable users by five upscaled grids.
More than 35\% of the users are still identifiable using the 1 km by 1 km grid as the original top four locations are still distinguishable.
5\% of the users are still identifiable using the 4 km by 4 km grid, which also made the observation area recognizable (Figure~\ref{fig:rescaled_result_4kv2}).

\input{top4_table}

\section*{Discussion}
\label{sec:discussion}

The observation area of the mobile positioning data sets is usually communicated.
However, a recently published data set \cite{yabe2023metropolitan}, did not disclose the geographic location of the mobility traces.
This study examines how effective this step is in terms of user privacy.
The main finding is that mobile positioning datasets describe the urban landscapes where the mobility took place, so the geographic location can be identified even if it is not disclosed.
In addition to the `YJMob100K' data set, this was demonstrated in four other cities from three other openly available data sets.
Furthermore, this effect has some resilience to upscaling, so that the urban area can be identified even with lower resolution grids.

It is important to emphasize that the goal of this work is not to demonstrate template matching on mobility data but to point out that human activity has a peculiar spatial distribution that makes the urban landscape recognizable even after discretization.
In cryptography, it is widely accepted that there is no security by obscurity \cite{diehl2016law3}.
Based on this principle, privacy through obscurity could not provide privacy.
However, in this case, the obscured observation area was meant to provide another layer of protection in addition to the spatial and temporal discretization.
Zhang and Bolot argue that publishing location data is likely to lead to privacy risks, and the data must be coarse in either the time domain or the space domain \cite{zang2011anonymization}.
Even partial location data can be used to infer the location.
In \cite{meteriz2022learning}, the altitude information from fitness tracker applications was used to infer the location of users.

This discourse leads to a trade-off between privacy preservation and researchers' interest in using more granular mobile positioning data to build better models.
In addition to adding noise to the location data \cite{acs2021privacy}, a possible solution for protecting user privacy when publishing mobility data is to exclude location information completely.
For example, a mobility data set was published \cite{du2018temporal}, covering Changchun Municipality, Northeast China, revealing only distances between locations.

This work is not without limitations.
First, the reconstructed grid cannot be arbitrarily accurate.
The accuracy of the grid anchor point is at most 500 meters in both directions (using the original resolution), because the template matching returns the coordinate of a pixel that represents a 500-meter by 500-meter cell.
As with the upscaled grids, the larger the grid size, the greater the chance that the anchor point will be misplaced.
The users' geographic locations are discretized into 500-meter by 500-meter grid cells, so the accuracy is inherently limited, but the inaccuracy of the reconstructed grid placement increases it further.

Note that the template matching was applied to the area surrounding the selected city with a limited size instead of the whole globe, which could be possible with sufficient computing capacity.
Furthermore, the applied template matching solution cannot deal with scaling or rotation (it was not necessary), although there are rotation and scale invariant template matching algorithms (e.g., \cite{amiri2010novel}).

Originally, the visibility of the road network in the heatmap \cite{yabe2023metropolitan} gave the idea that the urban landscape can be identified from mobility data, but using the road network with the presented template matching approach did not give satisfactory results.
This may be mainly because the roads are much narrower than the grid cells.
Also, the activity around the roads is not necessarily distributed proportionally to the priority or the order of a road.
For example, a highway is considered higher order than a street in a shopping district, but the latter may contain more user activity.
Thus, it is difficult to plot a map based solely on the cartographic features of the roads in a way that would be matched with the activity-weighted and spatially distorted templates.
For these reasons, the road network-based template matching could not match the landscape-based approach with the algorithm used.

The initial hypothesis of this study was that mobility trajectories can identify the urban area where the mobility took place, even if the location is not disclosed.
This hypothesis was tested on the `YJMob100K' data set, and is considered confirmed with the reconstructed grid.
The results show that hiding the observation area does not provide significant privacy benefits.
It was also shown that coarser grids can reveal the observation area.

\section*{Methods}
\label{sec:methods}

\begin{figure}[th]
    \begin{subfigure}[t]{0.245\linewidth}
        \includegraphics[width=\linewidth]{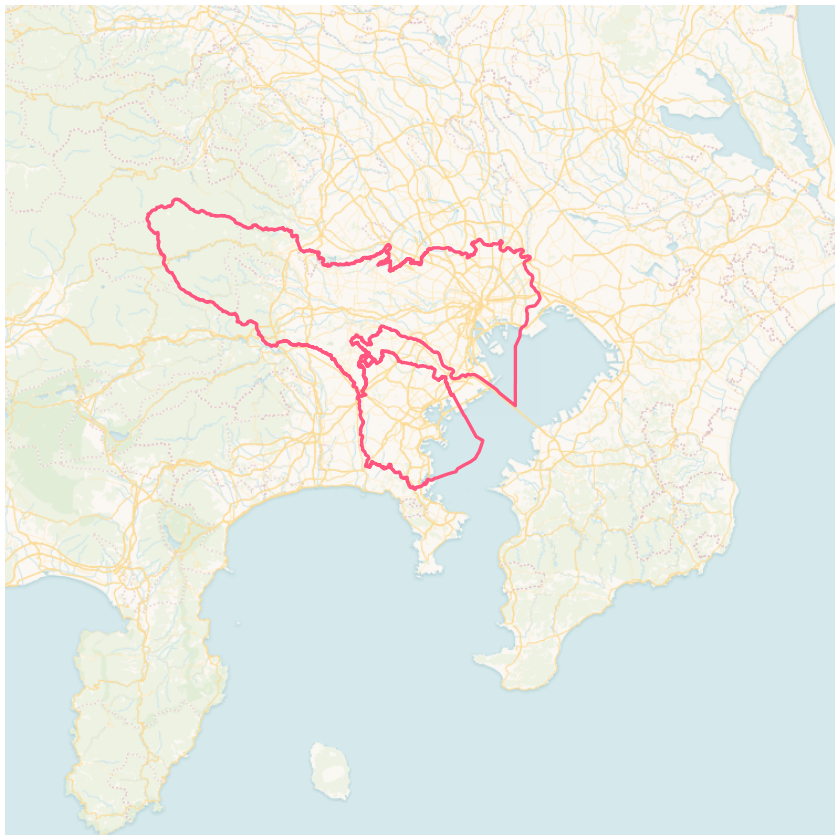}
        \captionsetup{position=bottom,justification=centering}
        \caption{}
        \label{fig:tokyo_and_yokohama}
    \end{subfigure}
    \hfill
    \begin{subfigure}[t]{0.245\linewidth}
        \includegraphics[width=\linewidth]{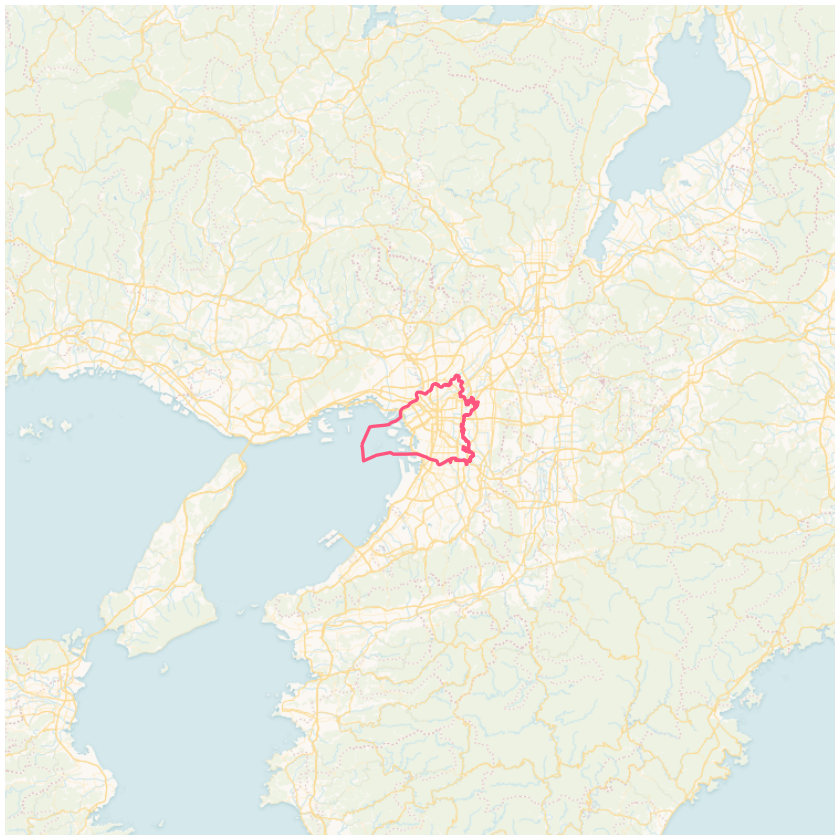}
        \captionsetup{position=bottom,justification=centering}
        \caption{}
        \label{fig:osaka}
    \end{subfigure}
    \hfill
    \begin{subfigure}[t]{0.245\linewidth}
        \includegraphics[width=\linewidth]{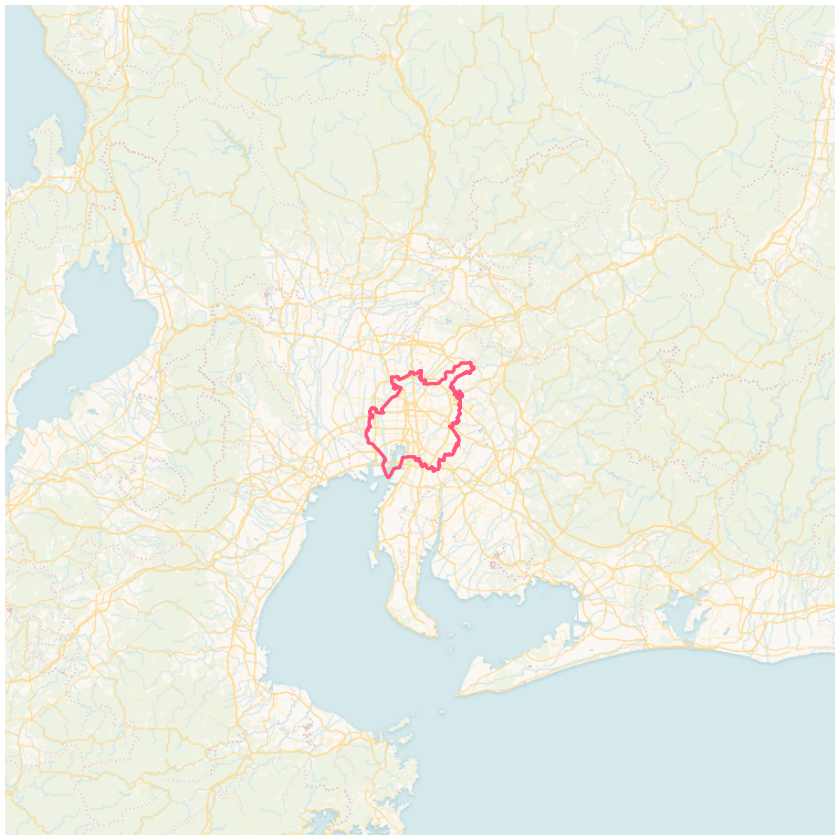}
        \captionsetup{position=bottom,justification=centering}
        \caption{}
        \label{fig:nagoya}
    \end{subfigure}
    \hfill
    \begin{subfigure}[t]{0.245\linewidth}
        \includegraphics[width=\linewidth]{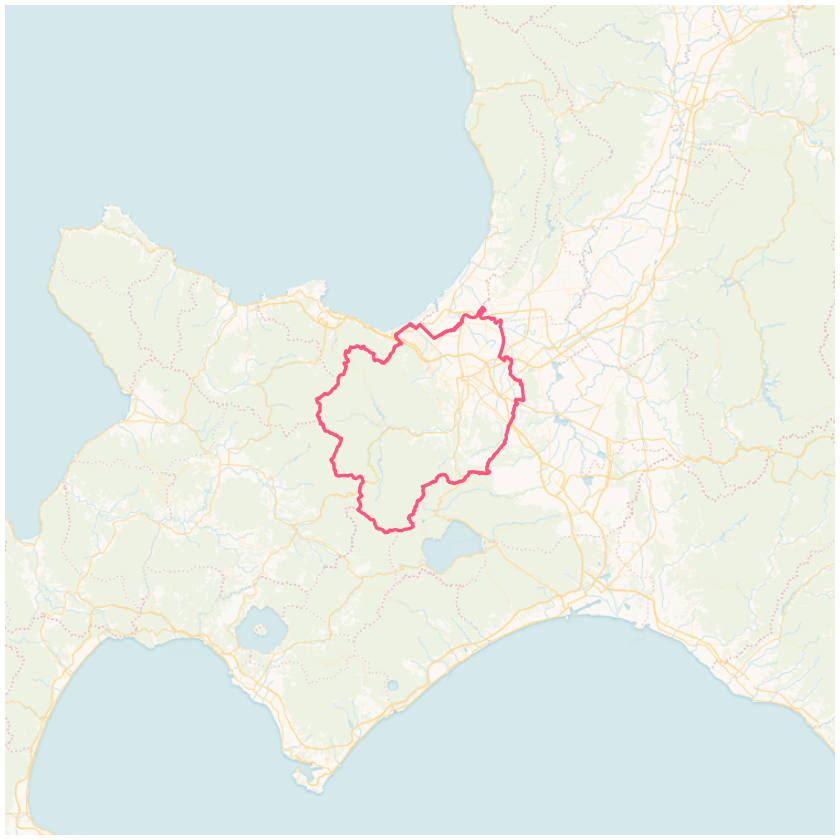}
        \captionsetup{position=bottom,justification=centering}
        \caption{}
        \label{fig:sapporo}
    \end{subfigure}
    \caption{
        The administrative boundaries of Japan's five largest cities: Tokyo (\textbf{\subref{fig:tokyo_and_yokohama}} top), Yokohama (\textbf{\subref{fig:tokyo_and_yokohama}} bottom), Osaka (\textbf{\subref{fig:osaka}}), Nagoya (\textbf{\subref{fig:nagoya}}) and Sapporo (\textbf{\subref{fig:sapporo}}).
    }
    \label{fig:largest_cities}
\end{figure}

Two pieces of information was required to deduce which city is in the focus of the data: (i) it was revealed that the metropolitan area is in Japan, and (ii) the urban area must be large as the observation area is 100 km x 100 km.
Taking into consideration the five largest cities of Japan\cite{enwiki2023largest} (Figure~\ref{fig:largest_cities}), the observation area might be Tokyo, Yokohama, Osaka, Nagoya or Sapporo, which are all near the sea (or ocean), so the low-activity parts might be waterfaces.


\begin{figure}[th]
    \centering
    \begin{subfigure}[t]{0.245\linewidth}
        \includegraphics[width=\linewidth]{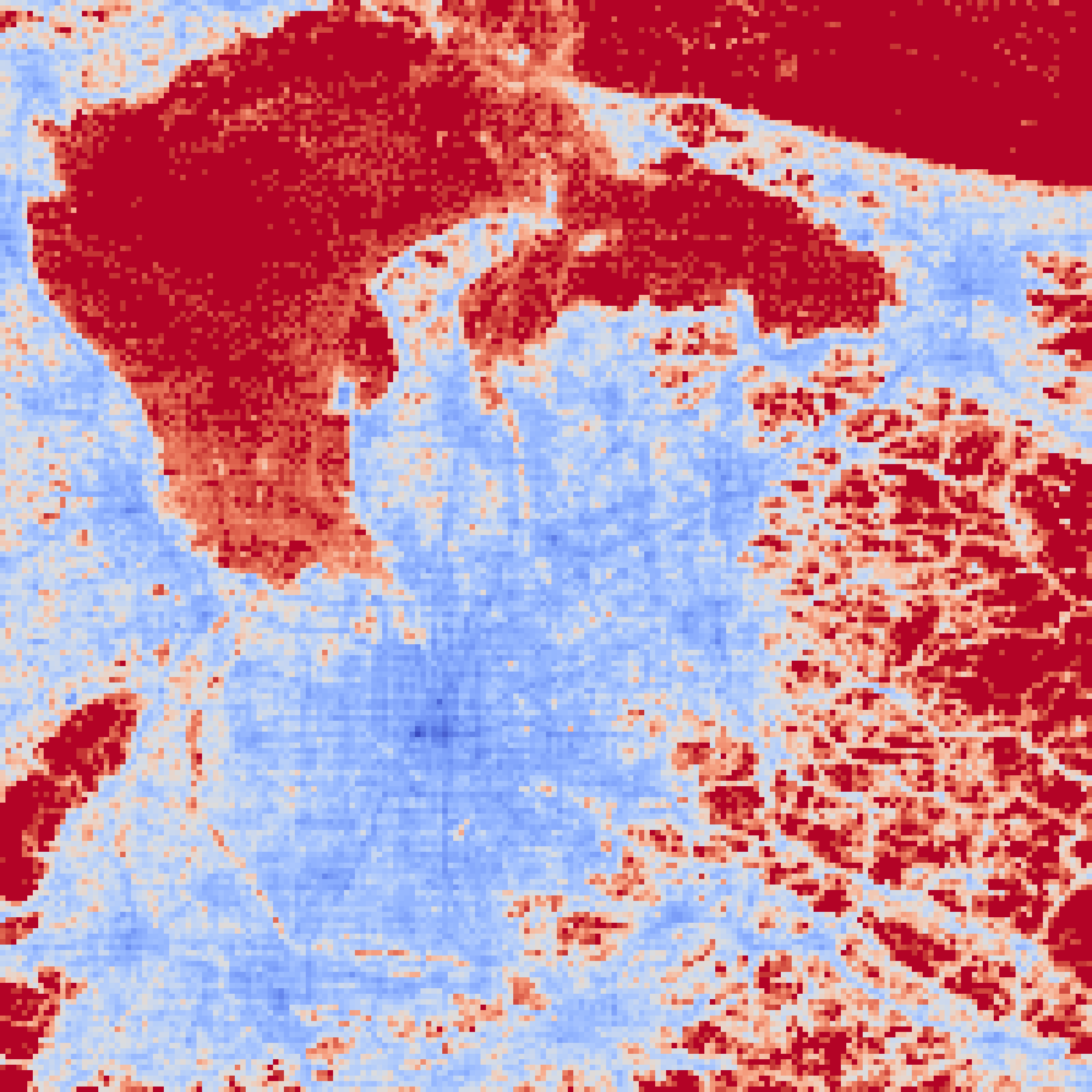}
        \captionsetup{position=bottom,justification=centering}
        \caption{}
        \label{fig:activity_heatmap}
    \end{subfigure}
    \begin{subfigure}[t]{0.245\linewidth}
        \includegraphics[width=\linewidth]{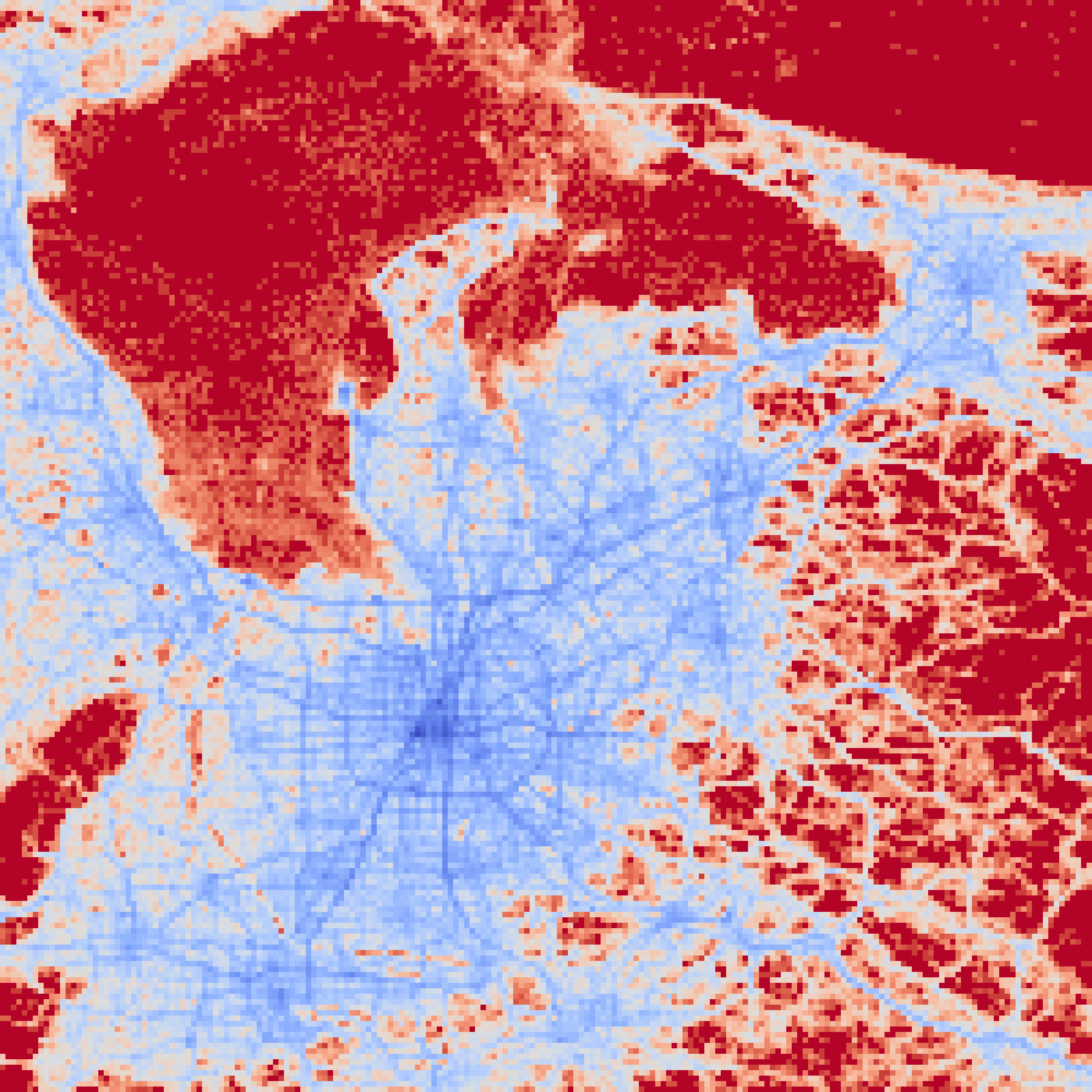}
        \captionsetup{position=bottom,justification=centering}
        \caption{}
        \label{fig:user_heatmap}
    \end{subfigure}
    \caption{
        Reproduction of \cite[Figure~6]{yabe2024yjmob100k} heatmaps, number of activity records (\textbf{\subref{fig:activity_heatmap}}) and unique users (\textbf{\subref{fig:user_heatmap}}) on log-scale.
    }
    \label{fig:reproduced_heatmaps}
\end{figure}

\begin{figure}[th]
    \centering
    \begin{subfigure}[t]{0.3\linewidth}
        \includegraphics[width=\linewidth]{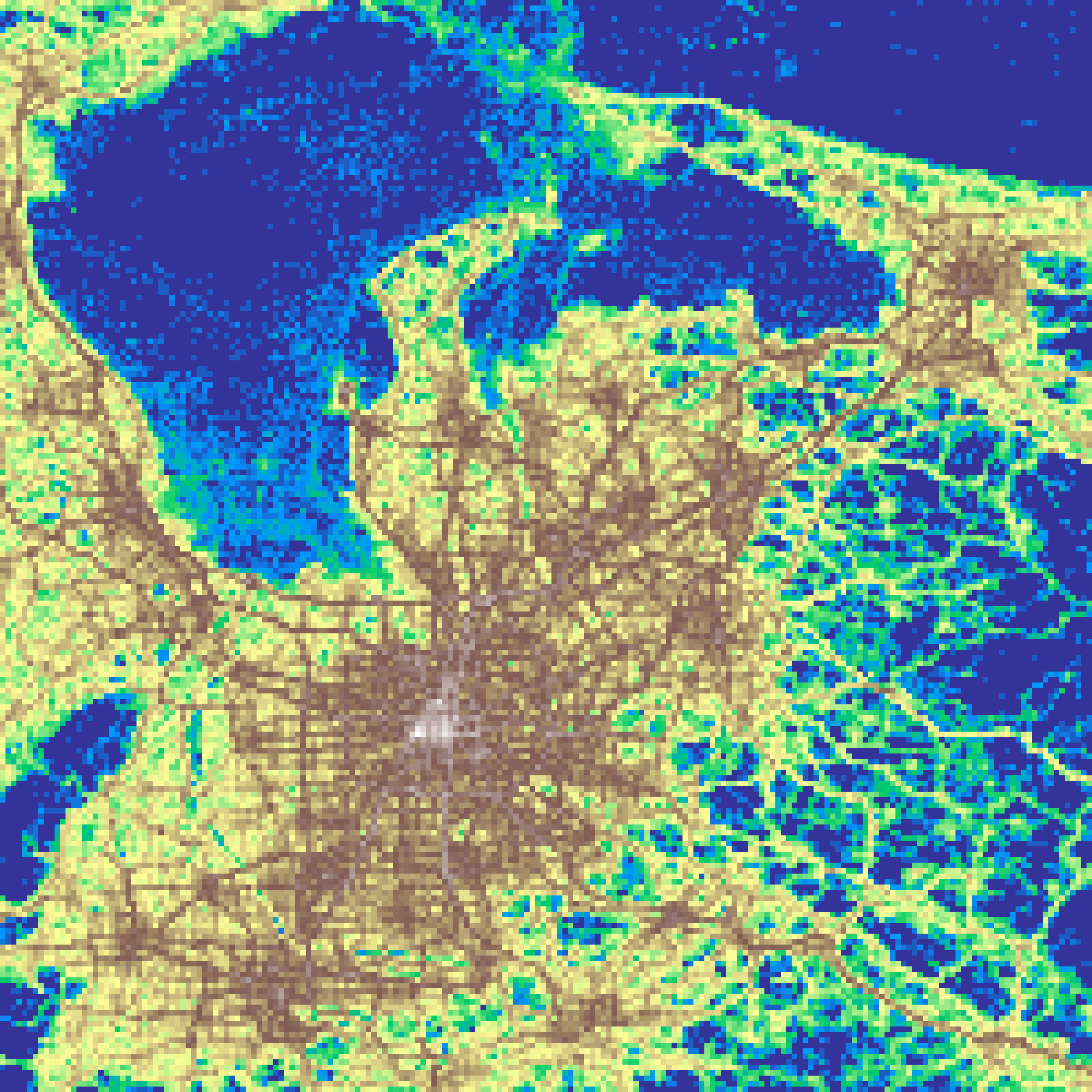}
        \captionsetup{position=bottom,justification=centering}
        \caption{}
        \label{fig:user_heatmap_terrain}
    \end{subfigure}
    \begin{subfigure}[t]{0.3\linewidth}
        \includegraphics[width=\linewidth]{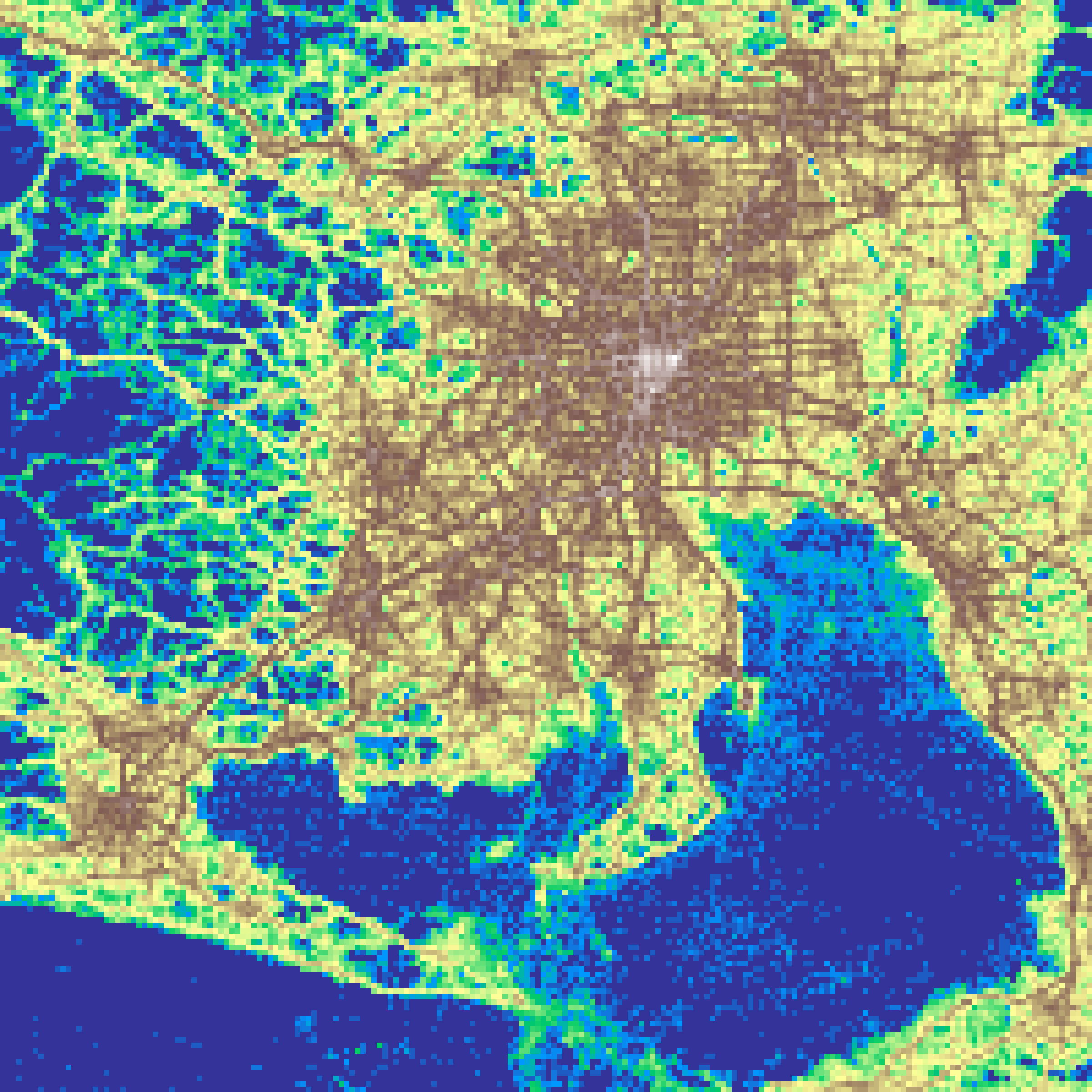}
        \captionsetup{position=bottom,justification=centering}
        \caption{}
        \label{fig:user_heatmap_terrain_rot}
    \end{subfigure}
    \begin{subfigure}[t]{0.3\linewidth}
        \includegraphics[width=\linewidth]{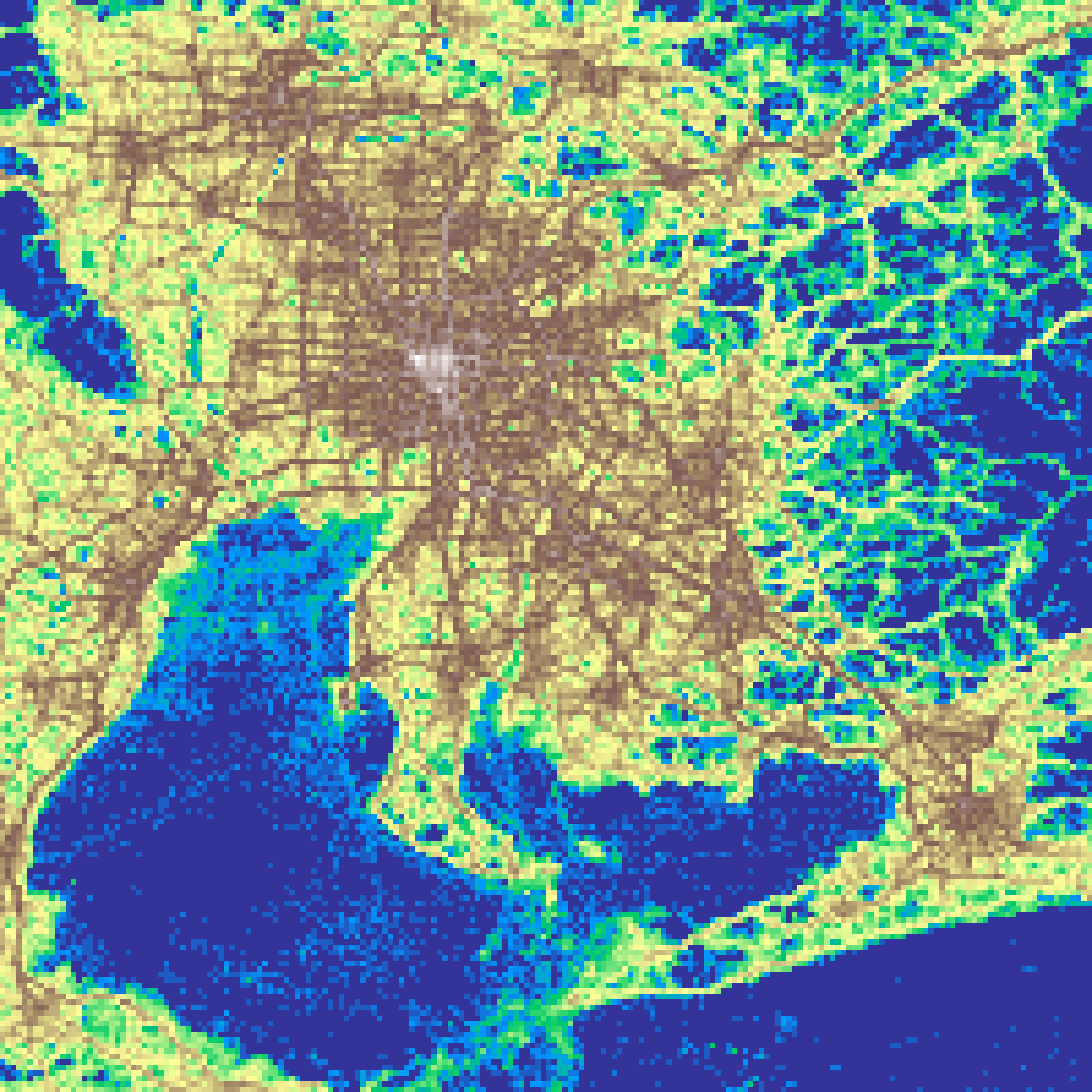}
        \captionsetup{position=bottom,justification=centering}
        \caption{}
        \label{fig:user_heatmap_terrain_fixed}
    \end{subfigure}
    \caption{
        The heatmaps showing the number of unique users using the `terrain' color palette (\textbf{\subref{fig:user_heatmap_terrain}}), rotated by \ang{180} (\textbf{\subref{fig:user_heatmap_terrain_rot}}) and flipped horizontally (\textbf{\subref{fig:user_heatmap_terrain_fixed}}).
    }
    \label{fig:transformations_part1}
\end{figure}

First, the Figure 6 of the data description paper \cite{yabe2024yjmob100k} was reproduced (Figure~\ref{fig:reproduced_heatmaps}).
It shows a 2-dimensional histogram of the number of pings and the number of observed unique users over the 75 days on a log-scale.
The urban area is clearly visible, especially in Figure~\ref{fig:user_heatmap}, where even the road network is visible.
Using a more natural color palette (Figure~\ref{fig:user_heatmap_terrain}) and comparing the heatmap with the cities from Figure~\ref{fig:largest_cities}, it is clear that the observation area is the Nagoya metropolitan area, but the figure is transformed.
After rotating the image by \ang{180} (Figure~\ref{fig:user_heatmap_terrain_rot}) and flipping it horizontally (Figure~\ref{fig:user_heatmap_terrain_fixed}) (or flipping it vertically), it becomes clear that the large red areas at the top of the original heatmaps (Figure~\ref{fig:reproduced_heatmaps}) are actually Ise Bay and Mikawa Bay.

\begin{figure}[th]
    \centering
    \begin{subfigure}[t]{0.245\linewidth}
        \includegraphics[width=\linewidth]{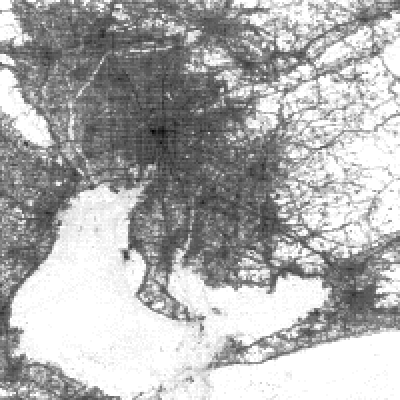}
        \captionsetup{position=bottom,justification=centering}
        \caption{}
        \label{fig:activity_greys}
    \end{subfigure}
    \begin{subfigure}[t]{0.345\linewidth}
        \includegraphics[width=\linewidth]{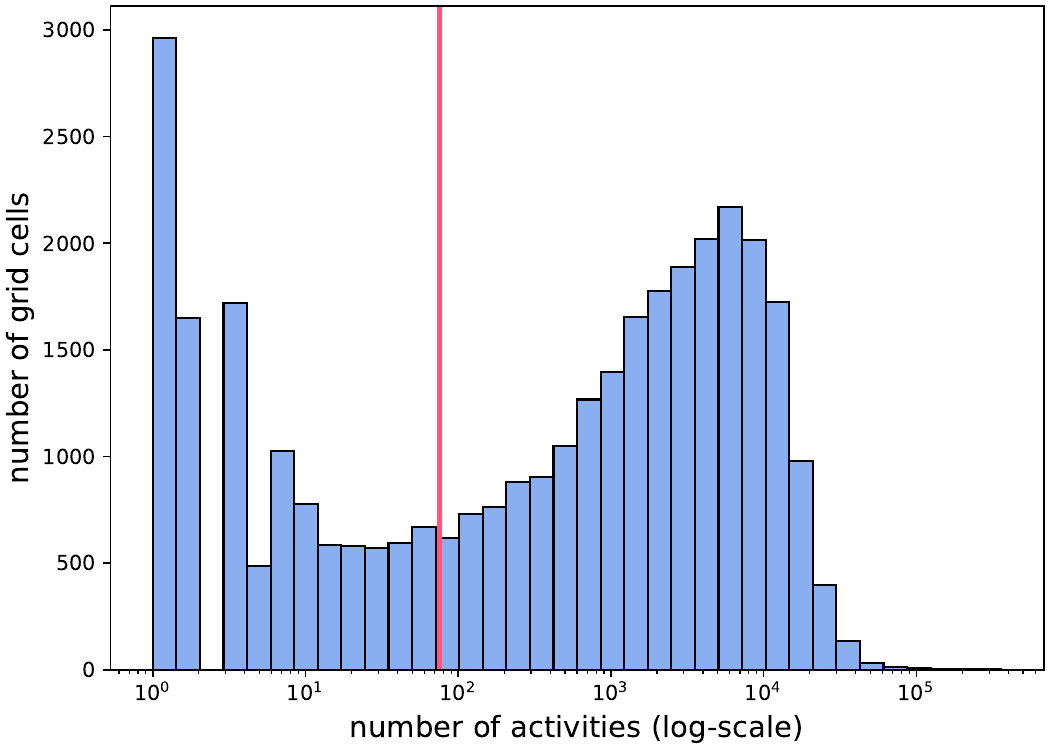}
        \captionsetup{position=bottom,justification=centering}
        \caption{}
        \label{fig:activity_histogram}
    \end{subfigure}
    \begin{subfigure}[t]{0.245\linewidth}
        \includegraphics[width=\linewidth]{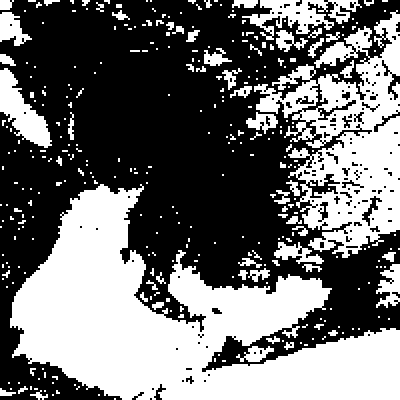}
        \captionsetup{position=bottom,justification=centering}
        \caption{}
        \label{fig:activity_cut}
    \end{subfigure}
    \caption{
        The thresholding process; starting from the grayscale heatmap of the activity (\textbf{\subref{fig:activity_greys}}), the histogram of the activity records (\textbf{\subref{fig:activity_histogram}}) with the selected threshold value (horizontal line), and the thresholded image (\textbf{\subref{fig:activity_cut}}).
    }
    \label{fig:thresholding}
\end{figure}

\begin{figure}[th!]
    \centering
    \begin{subfigure}[t]{0.245\linewidth}
        \includegraphics[width=\linewidth]{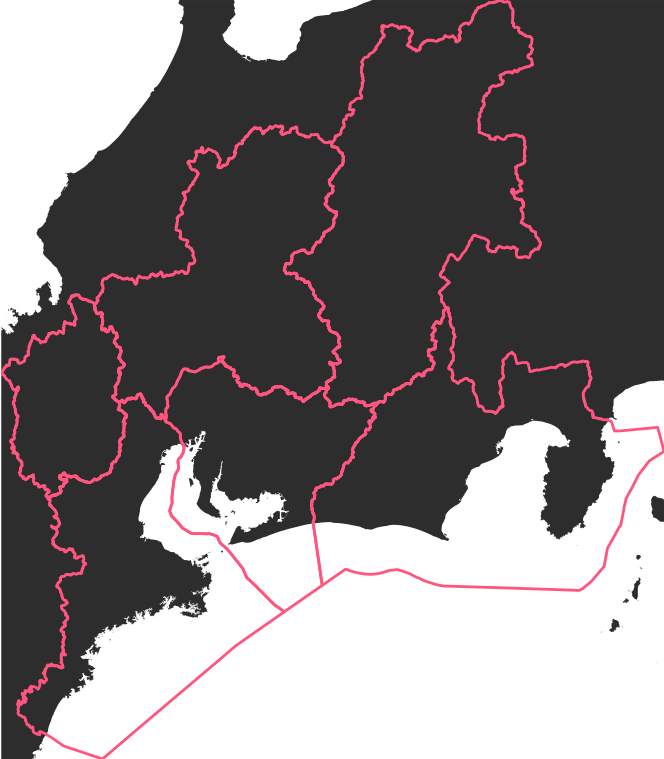}
        \captionsetup{position=bottom,justification=centering}
        \caption{}
        \label{fig:middle_japan}
    \end{subfigure}
    \begin{subfigure}[t]{0.245\linewidth}
        \includegraphics[width=\linewidth]{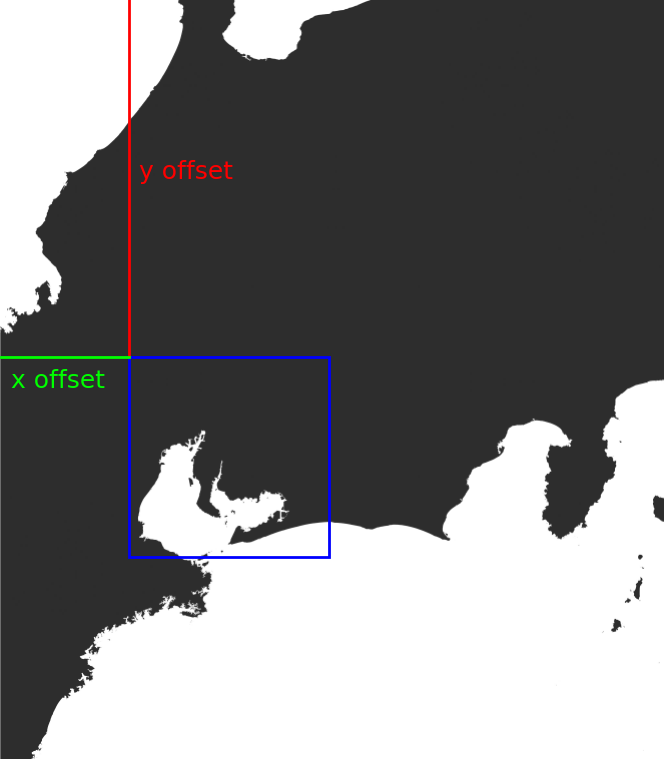}
        \captionsetup{position=bottom,justification=centering}
        \caption{}
        \label{fig:location_explanation}
    \end{subfigure}
    \begin{subfigure}[t]{0.245\linewidth}
        \includegraphics[width=\linewidth]{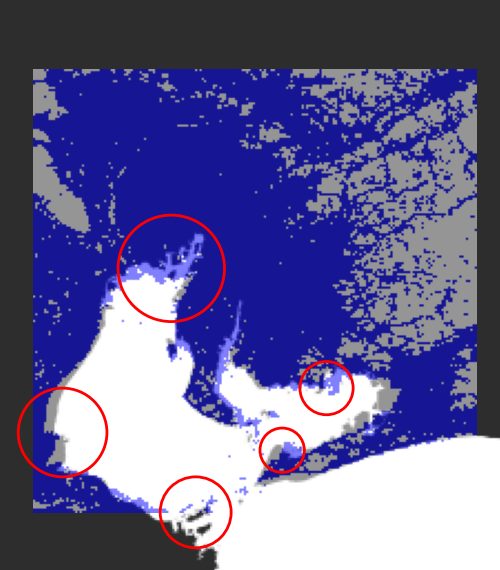}
        \captionsetup{position=bottom,justification=centering}
        \caption{}
        \label{fig:template_matched_zoomed_unstretched}
    \end{subfigure}
    \begin{subfigure}[t]{0.245\linewidth}
        \includegraphics[width=\linewidth]{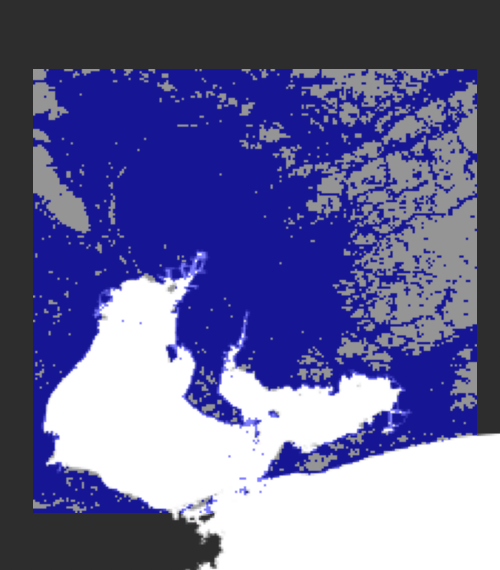}
        \captionsetup{position=bottom,justification=centering}
        \caption{}
        \label{fig:template_matched_zoomed}
    \end{subfigure}
    \caption{
        The land in the bounding box of the six selected prefectures (\textbf{\subref{fig:middle_japan}}), which serves as the input for the template matching (where Figure~\ref{fig:activity_cut} is the template), and the illustrated location of the result (\textbf{\subref{fig:location_explanation}}).
        The template is displayed over the map (\textbf{\subref{fig:template_matched_zoomed_unstretched}}), and some obvious mismatches are highlighted.
        After the correction (stretching) the overlapping shows perfect match (\textbf{\subref{fig:template_matched_zoomed}}).
    }
    \label{fig:locating_grid}
\end{figure}

The next problem is to find the exact location of the spatial grid that was used to discretize the spatial location of the mobility data.
The basic idea is to fit the heatmap to a proportional map as an image, taking advantage of the fact that Ise Bay and Mikawa Bay have a peculiar shape.
The plot of the log-scale activity (ping) heatmap using a grayscale palette is shown in Figure~\ref{fig:activity_greys}, Figure~\ref{fig:activity_histogram} shows the activity histogram, as well as the selected threshold (\num{75}).
Note that, since the threshold was applied directly to the data, the grayscale palette is not required, it is just shown for better understanding.
Figure~\ref{fig:activity_cut} shows the binary image as a result of the thresholding.
This process made the coastal area crisp by removing the activity at the ferry lines,  although there are low-activity rural lands at the western part of the region, that may appear as water in the binary image.

Six prefectures were downloaded from \acrfull*{OSM} via OSM-Boundaries \cite{osmboundaries}: Aichi, Gifu, Mie, Nagano, Shiga, Shizuoka.
The coastline of the Japanese islands was also needed, the data is from \acrshort*{OSM}, as pre-processed land polygons from \cite{osmcoastlines}, and the main islands of Japan were extracted manually.
Figure~\ref{fig:middle_japan} shows of a bounding box of the selected six prefectures, which will serve as the input for the template matching stage.
Template matching is a method to find the location of a template image in a larger image \cite{templatematching}.

The land geometry should be scaled to be proportional to the template, the observation area, which is a 200 by 200 grid.
The grid is plotted as a 200 by 200 pixel image, with each cell representing a 500 by 500 meter area.
Consequently, the bounding box of the selected six prefectures area should be plotted with a scale of 1 pixel = 500 meters.

Figure~\ref{fig:location_explanation} shows the result of the template matching, which determines the $x$ and the $y$ offset of the template image.
In this case, the thresholded activity heatmap (Figure~\ref{fig:activity_cut}).
The resulting area of the land (blue square in Figure~\ref{fig:location_explanation}) and its surroundings are shown in Figure~\ref{fig:template_matched_zoomed_unstretched} with the semitransparent template.


Figure~\ref{fig:template_matched_zoomed_unstretched} also highlights some obvious mismatches, although the matching is relatively correct.
It appears that the template has been squeezed vertically and horizontally stretched.
Since the template should not be transformed to maintain the conversion rate of `1 pixel equals to 500 meters', the land image was transformed instead.
To determine the exact horizontal and vertical stretching ratio, the background and the template were loaded into an image manipulation program, and the layer was manually stretched by trial and error.
The final values are 10\% horizontal stretching and 10\% vertical shrinking.
Figure~\ref{fig:template_matched_zoomed} shows the same comparison after these transformations were applied to the background image, and the match is perfect.

The template matching returns the $x$ and $y$ offsets of the template (Figure~\ref{fig:location_explanation}) in the input image, which was transformed and scaled to be proportional to the template.
Applying the inverse transformations to the offset coordinates determines the top-left point of the grid in the geographic coordinate system.
Using the reference point, a 200 by 200 grid should be generated with a 500-meter width and height.

The cell coordinates start from the upper left corner according to the description paper \cite{yabe2024yjmob100k} and increase to right and down.
There are two possible ways to join the challenge data with the reconstructed grid: (i) transform the coordinates in the mobility data and start coordinates as described, or (ii) assign transformed coordinate properties to the grid geometry and keep the mobility data as provided.
This work uses the second option. 
The transformed grid coordinates are visualized in Figure~\ref{fig:activity_count_log_on_map}.

\section*{Data availability}

The `YJMob100K' the `Weeplaces', and the Helsinki data sets are also available on Zenodo: \cite{yabe2024datasetv3}, \cite{chen2022weeplaces}, and \cite{tenkanen2019helsinki}, respectively.
The `Dallas--Fort Worth metroplex' data set is available from Dryad \cite{makris2023dfw}.

The census data was downloaded from the Portal Site of Official Statistics of Japan website (\href{https://www.e-stat.go.jp/}{https://www.e-stat.go.jp/}): Population, Households, Sex, Age and Marital Status, \textit{Population Census 2020}, Table 1-1.

\section*{Code availability}



The code to reproduce this work is available on GitHub: \href{https://github.com/pintergreg/reverse-engineering-YJMob100K-grid}{https://github.com/pintergreg/reverse-engineering-YJMob100K-grid}

\bibliography{references,deanonymization}


\section*{Acknowledgements} 


The map data is copyrighted by \acrlong*{OSM} contributors, available under \acrshort*{ODbL}, and the map tiles by Carto.



\section*{Competing interests} 

The author declares no competing interests.

\end{document}

%% file: acronyms.tex
\newacronym{BCR}{BCR}{Barrier Crossing Ratio}
\newacronym{D4D}{D4D}{Data for Development}
\newacronym{GIS}{GIS}{Geographic Information System}
\newacronym{GPS}{GPS}{Global Positioning System}
\newacronym{ECI}{ECI}{Economic Complexity Index}
\newacronym{HCSO}{HCSO}{Hungarian Central Statistical Office}
\newacronym{IMDb}{IMDb}{Internet Movie Database}
\newacronym{LBSN}{LBSN}{Location-Based Social Networks}
\newacronym{NYC}{NYC}{New York City}
\newacronym{ODbL}{ODbL}{Open Database License}
\newacronym{OLS}{OLS}{Ordinary Least Squares}
\newacronym{OSM}{OSM}{OpenStreetMap}
\newacronym{POI}{POI}{Point of Interest}
\newacronym{SAD}{SAD}{Symmetric Area Difference}
\newacronym{WGS84}{WGS84}{World Geodetic System}

%% file: population_correlations.tex
\newcommand{\NagoyaWardCorrelation}{0.8744}
\newcommand{\CityCorrelation}{0.8879}
\newcommand{\CityCorrelationWithoutKawagoe}{0.9717}

%% file: population_correlations_unstretched.tex
\newcommand{\NagoyaWardCorrelationUnstretched}{0.4676}
\newcommand{\CityCorrelationUnstretched}{0.7536}

%% file: toronto_stats.tex
\newcommand{\TorontoActivityCount}{110829}
\newcommand{\TorontoNumberOfUniqueUsers}{19356}
\newcommand{\NagoyaActivityCount}{111535175}

%% file: fix_wlscirep_sectionref.tex
\titleformat{name=\section,numberless}
  {\large\sffamily\bfseries}
  {}
  {0em}
  {\leavevmode\MakeLinkTarget[section]{}\ignorespaces#1}
  []  

\ExplSyntaxOn\makeatletter
\def\ttl@straight@i#1[#2]#3{%
  \tl_if_empty:nTF {#2}
   {\NR@gettitle{#3}}
   {\NR@gettitle{#2}}
  \gdef\ttl@savemark{\csname#1mark\endcsname{#3}}%
  \let\ttl@savewrite\@empty
  \def\ttl@savetitle{#3}%
  \gdef\thetitle{\csname the#1\endcsname}%
  \if@noskipsec \leavevmode \fi
  \par
  \ttl@labelling{#1}{#2}%
  \ttl@startargs\ttl@straight@ii{#1}{#3}}

\ExplSyntaxOff\makeatother

%% file: top4_table.tex
\begin{table}[t]
\centering
\caption{Comparison of the top-four-location identifiable users by upscaled grids.}
\label{tab:top4}
\begin{tabular}{rrrrrr}
\toprule
distinguishable cells & 1 km x 1 km & 2 km x 2 km & 4 km x 4 km & 8 km x 8 km & 16 km x 16 km \\
\midrule
4 & 35469 & 12882 & 5090 & 1810 & 470 \\
3 & 48228 & 42323 & 28457 & 16752 & 7438 \\
2 & 15582 & 38548 & 50987 & 52608 & 44939 \\
1 & 721 & 6247 & 15466 & 28830 & 47153 \\
\bottomrule
\end{tabular}
\end{table}

%% file: paper.bbl
\begin{thebibliography}{10}
\urlstyle{rm}
\expandafter\ifx\csname url\endcsname\relax
  \def\url#1{\texttt{#1}}\fi
\expandafter\ifx\csname urlprefix\endcsname\relax\def\urlprefix{URL }\fi
\expandafter\ifx\csname doiprefix\endcsname\relax\def\doiprefix{DOI: }\fi
\providecommand{\bibinfo}[2]{#2}
\providecommand{\eprint}[2][]{\url{#2}}

\bibitem{narayanan2008robust}
\bibinfo{author}{Narayanan, A.} \& \bibinfo{author}{Shmatikov, V.}
\newblock \bibinfo{title}{Robust de-anonymization of large sparse datasets}.
\newblock In \emph{\bibinfo{booktitle}{2008 IEEE Symposium on Security and Privacy (sp 2008)}}, \bibinfo{pages}{111--125} (\bibinfo{organization}{IEEE}, \bibinfo{year}{2008}).

\bibitem{douriez2016anonymizing}
\bibinfo{author}{Douriez, M.}, \bibinfo{author}{Doraiswamy, H.}, \bibinfo{author}{Freire, J.} \& \bibinfo{author}{Silva, C.~T.}
\newblock \bibinfo{title}{Anonymizing nyc taxi data: Does it matter?}
\newblock In \emph{\bibinfo{booktitle}{2016 IEEE international conference on data science and advanced analytics (DSAA)}}, \bibinfo{pages}{140--148} (\bibinfo{organization}{IEEE}, \bibinfo{year}{2016}).

\bibitem{tayouri2016social}
\bibinfo{author}{Tayouri, D.}
\newblock \bibinfo{journal}{\bibinfo{title}{Social media as an intelligence goldmine}}.
\newblock {\emph{\JournalTitle{Cyber security review}}} \bibinfo{pages}{27--30} (\bibinfo{year}{2016}).

\bibitem{lavrenovs2016privacy}
\bibinfo{author}{Lavrenovs, A.} \& \bibinfo{author}{Podins, K.}
\newblock \bibinfo{title}{Privacy violations in riga open data public transport system}.
\newblock In \emph{\bibinfo{booktitle}{2016 IEEE 4th Workshop on Advances in Information, Electronic and Electrical Engineering (AIEEE)}}, \bibinfo{pages}{1--6} (\bibinfo{organization}{IEEE}, \bibinfo{year}{2016}).

\bibitem{blondel2012data}
\bibinfo{author}{Blondel, V.~D.} \emph{et~al.}
\newblock \bibinfo{title}{Data for development: the d4d challenge on mobile phone data}, \url{10.48550/ARXIV.1210.0137} (\bibinfo{year}{2012}).

\bibitem{sharad2013anonymizing}
\bibinfo{author}{Sharad, K.} \& \bibinfo{author}{Danezis, G.}
\newblock \bibinfo{title}{De-anonymizing d4d datasets}.
\newblock In \emph{\bibinfo{booktitle}{Workshop on hot topics in privacy enhancing technologies}}, \bibinfo{pages}{10--10} (\bibinfo{organization}{Citeseer}, \bibinfo{year}{2013}).

\bibitem{de2013unique}
\bibinfo{author}{De~Montjoye, Y.-A.}, \bibinfo{author}{Hidalgo, C.~A.}, \bibinfo{author}{Verleysen, M.} \& \bibinfo{author}{Blondel, V.~D.}
\newblock \bibinfo{journal}{\bibinfo{title}{Unique in the crowd: The privacy bounds of human mobility}}.
\newblock {\emph{\JournalTitle{Scientific reports}}} \textbf{\bibinfo{volume}{3}}, \bibinfo{pages}{1--5} (\bibinfo{year}{2013}).

\bibitem{faraji2023point2hex}
\bibinfo{author}{Faraji, A.} \emph{et~al.}
\newblock \bibinfo{title}{Point2hex: Higher-order mobility flow data and resources}.
\newblock In \emph{\bibinfo{booktitle}{Proceedings of the 31st ACM International Conference on Advances in Geographic Information Systems}}, \bibinfo{pages}{1--4} (\bibinfo{year}{2023}).

\bibitem{bergroth202224}
\bibinfo{author}{Bergroth, C.}, \bibinfo{author}{J{\"a}rv, O.}, \bibinfo{author}{Tenkanen, H.}, \bibinfo{author}{Manninen, M.} \& \bibinfo{author}{Toivonen, T.}
\newblock \bibinfo{journal}{\bibinfo{title}{A 24-hour population distribution dataset based on mobile phone data from helsinki metropolitan area, finland}}.
\newblock {\emph{\JournalTitle{Scientific data}}} \textbf{\bibinfo{volume}{9}}, \bibinfo{pages}{39} (\bibinfo{year}{2022}).

\bibitem{xu2021towards}
\bibinfo{author}{Xu, Y.}, \bibinfo{author}{Xue, J.}, \bibinfo{author}{Park, S.} \& \bibinfo{author}{Yue, Y.}
\newblock \bibinfo{journal}{\bibinfo{title}{Towards a multidimensional view of tourist mobility patterns in cities: A mobile phone data perspective}}.
\newblock {\emph{\JournalTitle{Computers, Environment and urban systems}}} \textbf{\bibinfo{volume}{86}}, \bibinfo{pages}{101593} (\bibinfo{year}{2021}).

\bibitem{yabe2024yjmob100k}
\bibinfo{author}{Yabe, T.} \emph{et~al.}
\newblock \bibinfo{journal}{\bibinfo{title}{Yjmob100k: City-scale and longitudinal dataset of anonymized human mobility trajectories}}.
\newblock {\emph{\JournalTitle{Scientific Data}}} \textbf{\bibinfo{volume}{11}}, \bibinfo{pages}{397} (\bibinfo{year}{2024}).

\bibitem{yabe2024enhancing}
\bibinfo{author}{Yabe, T.} \emph{et~al.}
\newblock \bibinfo{journal}{\bibinfo{title}{Enhancing human mobility research with open and standardized datasets}}.
\newblock {\emph{\JournalTitle{Nature Computational Science}}} \bibinfo{pages}{1--4} (\bibinfo{year}{2024}).

\bibitem{acs2021privacy}
\bibinfo{author}{Acs, G.}, \bibinfo{author}{Lesty{\'a}n, S.} \& \bibinfo{author}{Bicz{\'o}k, G.}
\newblock \bibinfo{title}{Privacy of aggregated mobility data}.
\newblock In \emph{\bibinfo{booktitle}{Encyclopedia of Cryptography, Security and Privacy}}, \bibinfo{pages}{1--5} (\bibinfo{publisher}{Springer}, \bibinfo{year}{2021}).

\bibitem{mir2013dp}
\bibinfo{author}{Mir, D.~J.}, \bibinfo{author}{Isaacman, S.}, \bibinfo{author}{C{\'a}ceres, R.}, \bibinfo{author}{Martonosi, M.} \& \bibinfo{author}{Wright, R.~N.}
\newblock \bibinfo{title}{Dp-where: Differentially private modeling of human mobility}.
\newblock In \emph{\bibinfo{booktitle}{2013 IEEE international conference on big data}}, \bibinfo{pages}{580--588} (\bibinfo{organization}{IEEE}, \bibinfo{year}{2013}).

\bibitem{juhasz2023amenity}
\bibinfo{author}{Juh{\'a}sz, S.} \emph{et~al.}
\newblock \bibinfo{journal}{\bibinfo{title}{Amenity complexity and urban locations of socio-economic mixing}}.
\newblock {\emph{\JournalTitle{EPJ Data Science}}} \textbf{\bibinfo{volume}{12}}, \bibinfo{pages}{34} (\bibinfo{year}{2023}).

\bibitem{pappalardo2021evaluation}
\bibinfo{author}{Pappalardo, L.}, \bibinfo{author}{Ferres, L.}, \bibinfo{author}{Sacasa, M.}, \bibinfo{author}{Cattuto, C.} \& \bibinfo{author}{Bravo, L.}
\newblock \bibinfo{journal}{\bibinfo{title}{Evaluation of home detection algorithms on mobile phone data using individual-level ground truth}}.
\newblock {\emph{\JournalTitle{EPJ data science}}} \textbf{\bibinfo{volume}{10}}, \bibinfo{pages}{29} (\bibinfo{year}{2021}).

\bibitem{pinter2022awakening}
\bibinfo{author}{Pint{\'e}r, G.} \& \bibinfo{author}{Felde, I.}
\newblock \bibinfo{journal}{\bibinfo{title}{Awakening city: Traces of the circadian rhythm within the mobile phone network data}}.
\newblock {\emph{\JournalTitle{Information}}} \textbf{\bibinfo{volume}{13}}, \bibinfo{pages}{114} (\bibinfo{year}{2022}).

\bibitem{vanhoof2018assessing}
\bibinfo{author}{Vanhoof, M.}, \bibinfo{author}{Reis, F.}, \bibinfo{author}{Ploetz, T.} \& \bibinfo{author}{Smoreda, Z.}
\newblock \bibinfo{journal}{\bibinfo{title}{Assessing the quality of home detection from mobile phone data for official statistics}}.
\newblock {\emph{\JournalTitle{Journal of official statistics}}} \textbf{\bibinfo{volume}{34}}, \bibinfo{pages}{935--960} (\bibinfo{year}{2018}).

\bibitem{pinter2022commuting}
\bibinfo{author}{Pint{\'e}r, G.} \& \bibinfo{author}{Felde, I.}
\newblock \bibinfo{journal}{\bibinfo{title}{Commuting analysis of the budapest metropolitan area using mobile network data}}.
\newblock {\emph{\JournalTitle{ISPRS International Journal of Geo-Information}}} \textbf{\bibinfo{volume}{11}}, \bibinfo{pages}{466}, \url{10.3390/ijgi11090466} (\bibinfo{year}{2022}).

\bibitem{estat2020population}
\bibinfo{author}{{{Official Statistics of Japan}}}.
\newblock \bibinfo{title}{Population census 2020}.
\newblock \bibinfo{howpublished}{\url{https://www.e-stat.go.jp/en/stat-search/files?page=1&layout=datalist&toukei=00200521&tstat=000001136464&cycle=0&year=20200&month=24101210&tclass1=000001136466}} (\bibinfo{year}{2020}).
\newblock \bibinfo{note}{[Online; accessed 28-February-2024]}.

\bibitem{chen2022weeplaces}
\bibinfo{author}{Chen, Z.}
\newblock \bibinfo{title}{Spatiotemporal checkins with social connections}, \url{https://doi.org/10.5281/zenodo.6369319} (\bibinfo{year}{2022}).

\bibitem{chen2022contrasting}
\bibinfo{author}{Chen, Z.} \emph{et~al.}
\newblock \bibinfo{journal}{\bibinfo{title}{Contrasting social and non-social sources of predictability in human mobility}}.
\newblock {\emph{\JournalTitle{Nature Communications}}} \textbf{\bibinfo{volume}{13}}, \url{10.1038/s41467-022-29592-y} (\bibinfo{year}{2022}).

\bibitem{yabe2023dataset}
\bibinfo{author}{Yabe, T.} \emph{et~al.}
\newblock \bibinfo{title}{{YJMob100K: City-Scale and Longitudinal Dataset of Anonymized Human Mobility Trajectories}}, \url{10.5281/zenodo.10142719} (\bibinfo{year}{2023}).

\bibitem{makris2023dfw}
\bibinfo{author}{Makris, N.}, \bibinfo{author}{Moghimi, R.}, \bibinfo{author}{Godat, E.} \& \bibinfo{author}{Vu, T.}
\newblock \bibinfo{title}{Gps data from cellphones for mechanical analog}, \url{https://doi.org/10.5061/dryad.1c59zw3zg} (\bibinfo{year}{2023}).

\bibitem{makris2023mechanical}
\bibinfo{author}{Makris, N.}, \bibinfo{author}{Moghimi, G.}, \bibinfo{author}{Godat, E.} \& \bibinfo{author}{Vu, T.}
\newblock \bibinfo{journal}{\bibinfo{title}{Mechanical analogue for cities}}.
\newblock {\emph{\JournalTitle{Royal Society Open Science}}} \textbf{\bibinfo{volume}{10}}, \bibinfo{pages}{220943} (\bibinfo{year}{2023}).

\bibitem{h3}
\bibinfo{author}{{Uber Technologies, Inc.}}
\newblock \bibinfo{title}{H3} (\bibinfo{year}{2024}).
\newblock \bibinfo{note}{(Online; accessed on 3 May 2024)}.

\bibitem{zang2011anonymization}
\bibinfo{author}{Zang, H.} \& \bibinfo{author}{Bolot, J.}
\newblock \bibinfo{title}{Anonymization of location data does not work: A large-scale measurement study}.
\newblock In \emph{\bibinfo{booktitle}{Proceedings of the 17th annual international conference on Mobile computing and networking}}, \bibinfo{pages}{145--156} (\bibinfo{year}{2011}).

\bibitem{yabe2023metropolitan}
\bibinfo{author}{Yabe, T.} \emph{et~al.}
\newblock \bibinfo{title}{Metropolitan scale and longitudinal dataset of anonymized human mobility trajectories} (\bibinfo{year}{2023}).
\newblock \eprint{2307.03401}.

\bibitem{diehl2016law3}
\bibinfo{author}{Diehl, E.}
\newblock \emph{\bibinfo{title}{Law 3: No Security Through Obscurity}}, \bibinfo{pages}{67--79} (\bibinfo{publisher}{Springer International Publishing}, \bibinfo{address}{Cham}, \bibinfo{year}{2016}).

\bibitem{meteriz2022learning}
\bibinfo{author}{Meteriz-Yildiran, U.}, \bibinfo{author}{Yildiran, N.~F.}, \bibinfo{author}{Kim, J.} \& \bibinfo{author}{Mohaisen, D.}
\newblock \bibinfo{journal}{\bibinfo{title}{Learning location from shared elevation profiles in fitness apps: A privacy perspective}}.
\newblock {\emph{\JournalTitle{IEEE Transactions on Mobile Computing}}}  (\bibinfo{year}{2022}).

\bibitem{du2018temporal}
\bibinfo{author}{Du, Z.} \emph{et~al.}
\newblock \bibinfo{journal}{\bibinfo{title}{The temporal network of mobile phone users in changchun municipality, northeast china}}.
\newblock {\emph{\JournalTitle{Scientific data}}} \textbf{\bibinfo{volume}{5}}, \bibinfo{pages}{1--7} (\bibinfo{year}{2018}).

\bibitem{amiri2010novel}
\bibinfo{author}{Amiri, M.} \& \bibinfo{author}{Rabiee, H.~R.}
\newblock \bibinfo{journal}{\bibinfo{title}{A novel rotation/scale invariant template matching algorithm using weighted adaptive lifting scheme transform}}.
\newblock {\emph{\JournalTitle{Pattern Recognition}}} \textbf{\bibinfo{volume}{43}}, \bibinfo{pages}{2485--2496} (\bibinfo{year}{2010}).

\bibitem{enwiki2023largest}
\bibinfo{author}{{Wikipedia contributors}}.
\newblock \bibinfo{title}{Largest cities in japan by population by decade --- {Wikipedia}{,} the free encyclopedia} (\bibinfo{year}{2023}).
\newblock \bibinfo{note}{[Online; accessed 7 February 2024]}.

\bibitem{osmboundaries}
\bibinfo{title}{{OSM-Boundaries}}.
\newblock \bibinfo{howpublished}{\url{https://osm-boundaries.com/}} (\bibinfo{year}{2020-2024}).
\newblock \bibinfo{note}{[Online; accessed 8 February 2024]}.

\bibitem{osmcoastlines}
\bibinfo{title}{{Land polygons}}.
\newblock \bibinfo{howpublished}{\url{https://osmdata.openstreetmap.de/data/land-polygons.html}}.
\newblock \bibinfo{note}{[Online; accessed 8 February 2024]}.

\bibitem{templatematching}
\bibinfo{title}{{OpenCV: Template Matching}}.
\newblock \bibinfo{howpublished}{\url{https://docs.opencv.org/3.4/d4/dc6/tutorial_py_template_matching.html}}.
\newblock \bibinfo{note}{[Online; accessed 20 February 2024]}.

\bibitem{yabe2024datasetv3}
\bibinfo{author}{Yabe, T.} \emph{et~al.}
\newblock \bibinfo{title}{{YJMob100K: City-Scale and Longitudinal Dataset of Anonymized Human Mobility Trajectories}}, \url{https://doi.org/10.5281/zenodo.10836269} (\bibinfo{year}{2024}).

\bibitem{tenkanen2019helsinki}
\bibinfo{author}{Tenkanen, H.} \& \bibinfo{author}{Toivonen, T.}
\newblock \bibinfo{title}{Helsinki region travel time matrix}, \url{https://doi.org/10.5281/zenodo.3247564} (\bibinfo{year}{2019}).

\end{thebibliography}
